\documentclass[a4paper,11pt]{article}
\usepackage{jheppub} 
\usepackage{lineno}
\usepackage{macros}
\usepackage{theorems}
\usepackage{subcaption}
\usepackage{amsthm}

\newtheorem{innercustomthm}{Conjecture}
\newenvironment{customconj}[1]
  {\renewcommand\theinnercustomthm{#1}\innercustomthm}
  {\endinnercustomthm}

\newcommand{\sr}[1]{\color{purple}\textbf{#1}\textbf{ \footnotesize{-SR}}\color{black}}

\title{Swampland and the Geometry of Marked Moduli Spaces}

\author[a]{Sanjay Raman}
\author[a]{Cumrun Vafa}

\emailAdd{sanjayraman@g.harvard.edu}
\emailAdd{vafa@g.harvard.edu}

\affiliation[a]{Jefferson Physical Laboratory, Harvard University\\ 17 Oxford St, Cambridge, MA 02138, United States of America} 

\abstract{We define the notion of a {\it marked moduli space} as the parameter space of a physical theory together with all of its observables. In geometric examples, this coincides with the mathematical notion of Teichmüller space. We propose two new Swampland principles about the geometry of marked moduli spaces: We conjecture that a marked moduli space is always contractible, and moreover, that there is a unique shortest path connecting any pair of points in it with respect to its physical metric. We provide strong evidence for these conjectures for theories with 8 or more supercharges.}

\begin{document}
\maketitle
\flushbottom

\section{Introduction} \label{sec:intro}

The Swampland program \cite{vafa_string_2005, brennan_string_2018, van_beest_lectures_2022,agmon_lectures_2023,palti_swampland_2019,grana_swampland_2021} seeks to characterize the space of effective field theories (EFTs) coupled to gravity with a valid UV completion. A key approach to studying such theories is to constrain their field content and couplings. In this context, it is widely believed that every coupling in quantum gravity has its value set dynamically by the vacuum expectation value (vev) of a field \cite{ooguri_geometry_2007}. When these fields are massless scalars (as is the case in examples with enough supersymmetry), their vevs parametrize the exact moduli space of vacua of the theory, whose metric is specified by the coefficients of the scalars' kinetic term. These moduli spaces are therefore a central object of study. 

In \cite{ooguri_geometry_2007}, several conjectures were made about the geometry of such moduli spaces. Perhaps the most well-known of these is the \textit{distance (duality) conjecture}, which predicts an infinite tower of light particles at infinite field distance in the moduli space. It was also conjectured in \cite{ooguri_geometry_2007} that the scalar curvature of moduli spaces is negative in asymptotic regions. This conjecture was motivated by many examples from the string landscape: It is supported by a wealth of examples of 5d $\C{N}=1$ supergravity moduli spaces arising from the compactification of M-theory on a Calabi-Yau 3-fold (CY3). 
Nevertheless, even within the arena of CY3 compactifications of M-theory, it turns out that the prediction of asymptotic negative curvature is false \cite{trenner_asymptotic_2010, marchesano_moduli_2023}: Not only can the curvature be positive in asymptotic limits, it is shown to diverge to $+\infty$ in explicit examples \cite{trenner_asymptotic_2010}. The following question therefore remains: Is there any reformulation of the asymptotic negative curvature conjecture that is correct? One aim of this paper is to provide such a conjecture.

The canonical physical metric placed on moduli space arises from the coefficients of the kinetic term in the EFT action for the massless scalars. In the context of compactification of type IIB on CY3, where the complex structure moduli space corresponds to the physical vector multiplet moduli space in the effective theory, the physical metric is known as the \textit{Weil-Petersson} (WP) metric. However, it was observed first in the mathematics literature \cite{lu_hodge_2005, lu_curvature_2005} and later in the physics literature \cite{cecotti_moduli_2020, cecotti_special_2020, cecotti_swampland_2021} that the physical WP metric is not the only natural metric that can be placed on the moduli space. Indeed, a second metric, the \textit{Hodge metric}, arises by pulling back the homogeneous metric on the CY3 space of periods $D$ along the map $\C{M} \to \Gamma\backslash D$ (here $\Gamma$ is the monodromy group of the CY3 equipped with its natural action on the periods). The Hodge metric, in some sense, has better curvature properties than the WP metric --- it has nonpositive holomorphic bisectional curvature everywhere \cite{lu_curvature_2005}. However, since the Hodge metric is not the physical metric seen from the EFT viewpoint,\footnote{An attempt at its physical description was made by \cite{cecotti_moduli_2020} in terms of a (backwards) Ricci flow of the WP metric, but the Hodge metric resists a natural physical interpretation from the EFT perspective.} it is important to understand exactly which topological and geometric properties related to the negative Hodge curvature survive when we pass to the physical WP geometry. 

For the 5d $\C{N}=1$ case of M-theory on CY3, it was observed in \cite{marchesano_moduli_2023} that the limits with divergent positive curvature coincided with the emergence of a QFT sector decoupled from gravity living below the EFT cutoff scale set by the distance conjecture. This aligns with earlier observations in \cite{ooguri_geometry_2007} that singular points at \textit{finite} distance in the CY3 moduli space with divergent positive curvature are related to the appearance of a decoupled QFT in the corresponding singular CY3 geometry, which is why the original conjecture predicted negative curvature only in the asymptotic regions of the moduli space. Therefore, in the 4d case of IIB on CY3, it is also important to understand the behavior of the WP geometry in the vicinity of these finite-distance QFT points. 

Another set of conjectures in \cite{ooguri_geometry_2007} is related to the topology of moduli spaces. In particular, it was conjectured that there is no non-trivial 1-cycle with minimum length within a given homotopy class in moduli space. It turns out, however, that an analogous statement cannot be carried over to higher homotopy classes. For example, the second homotopy group of the moduli space of complex structures of $T^2$ (realized, for instance, as the axiodilaton moduli space of the F-theory torus) is not trivial, and the minimum area of nontrivial 2-cycles is positive. 

In this paper, we will propose a new Swampland conjecture which captures and generalizes the spirit of both the asymptotic negative curvature conjecture and the trivial fundamental group conjecture. It is our aim to state and provide evidence for this conjecture in this paper. To do this, we introduce a new notion of moduli space, the \textit{marked moduli space}, which is defined to be the moduli space of a physical theory along with a choice of basis for its observables. This terminology is similar to what one uses when one talks about the moduli space of \textit{marked Riemann surfaces}, which are Riemann surfaces equipped with a specific choice of a class of 1-cycles. Such a moduli space with marking becomes, in geometric examples, what is known in the mathematical literature as \textit{Teichmüller space}. For example, in the case of $T^2$, the Teichmüller space is precisely the upper half-plane $\B{H}$, \textit{without} the subsequent quotient by the action of the modular group $\Gamma = \PSL(2, \B{Z})$. In the context of string theory (for instance, type IIB on $T^2$), a marking corresponds to a choice of how to label observable classes (such as branes wrapping specific cycles).

Our main conjectures are the following two related statements about the geometry and topology of marked moduli spaces. In what follows, let $\hat{\C{M}}$ denote the marked moduli space of an EFT coupled to gravity. Then we conjecture the following: 

\begin{customconj}{1} \label{conj1}
    The space $\hat{\C{M}}$ is contractible.
\end{customconj} 

\begin{customconj}{2'} \label{conj2'}
    The asymptotic regions of the space $\hat{\C{M}}$ have the following property with respect to the physical metric: For each pair of points $p, q$ at asymptotically infinite distance from the interior of $\hat{\C{M}}$, there is at most one geodesic between them. That is, if $p, q$ are connected by a geodesic, that geodesic is unique.
\end{customconj}

Conjecture \ref{conj1} is a global topological statement about marked moduli spaces in general, while Conjecture \ref{conj2'} pertains to the behavior of geodesics only in asymptotic regions of marked moduli spaces approaching infinite distance from the moduli space interior. In effect, Conjecture \ref{conj2'} replaces the asymptotic negative curvature conjecture of \cite{ooguri_geometry_2007}. Given these statements, it is natural to conjecture a much stronger statement that implies both \ref{conj1} and \ref{conj2'}: 

\begin{customconj}{2} \label{conj2}
    The marked moduli space $\hat{\C{M}}$ satisfies what we will call the \emph{unique shortest path property}. This amounts to the following pair of statements: 
    \begin{enumerate}
        \item There is at most one geodesic between any pair of points in $\hat{\C{M}}$. 
        \item There is exactly one shortest path between any pair of points in $\hat{\C{M}}$, and this path is piecewise geodesic, possibly connecting through singular points of $\hat{\C{M}}$ which lie at finite distance.
    \end{enumerate} 
\end{customconj}

In the case where $\hat{\C{M}}$ is complete and there are no finite-distance curvature singularities, the statement of Conjecture \ref{conj2} is equivalent to the \textit{unique geodesic property}: every pair of points in $\hat{\C{M}}$ is connected by a unique geodesic. Moreover, the unique geodesic property implies Conjecture \ref{conj2}, so to establish Conjecture \ref{conj2} in any given example, it is always enough to show the unique geodesic property. In this sense, Conjecture \ref{conj2} is a technical refinement of the unique geodesic property in marked moduli spaces with finite-distance curvature singularities, and should be viewed as preserving the ``spirit'' of the unique geodesic property in more general singular spaces. We will see examples where this technical condition becomes relevant in Section \ref{sec:cft}.

Note also that neither Conjecture \ref{conj1} nor Conjecture \ref{conj2'} implies the other, but Conjecture \ref{conj2} implies both Conjectures \ref{conj1} (through contraction along shortest paths) and \ref{conj2'} (as it generalizes it). We do not know of any counterexamples to Conjecture \ref{conj2}, and we provide evidence towards it in Section \ref{sec:cft}, but nevertheless Conjecture \ref{conj2} remains a more speculative statement.  The main reason to believe Conjecture \ref{conj2} is its simplicity and the fact that it makes a general statement about quantum gravity (and not just its asymptotic regime).

The sense in which these conjectures modify and combine existing Swampland conjectures in \cite{ooguri_geometry_2007} is as follows. In various examples, the contractibility of a space is deeply related to the nonpositivity of its sectional curvature, together with triviality of the fundamental group. This proceeds by establishing a unique geodesic between any pair of points, which is a (stronger version of) Conjecture \ref{conj2}. Our view here is that the asymptotic negativity of the curvature proposed in \cite{ooguri_geometry_2007} should be considered as aiming to argue for the contractibility and the uniqueness of shortest paths between pairs of points in the marked moduli space, at least asymptotically. The conjectures we propose here achieve the same objective.

In theories with $Q > 8$ supercharges, the contractibility of marked moduli spaces can be shown directly from the uniqueness of geodesics in the physical metric, which in turn is related to nonpositivity bounds on the sectional curvature.  For $Q = 8$ supercharges, it is not true anymore that moduli spaces have nonpositive curvature in the physical metric, but we will nevertheless prove Conjecture \ref{conj1} and provide strong evidence for Conjecture \ref{conj2'} in these cases as well.\footnote{For less supersymmetry, it becomes difficult to discuss the local (and therefore, the global) geometry of moduli spaces, since a superpotential can arise. We defer the discussion of such cases to future work.} Note also that our Conjecture \ref{conj2} pertains to the \textit{global} geometry of the moduli space and is not restricted in scope to the asymptotic regime, like our Conjecture \ref{conj2'} or the negative curvature conjecture of \cite{ooguri_geometry_2007}.\footnote{It is possible that our second conjecture is valid only in its asymptotic form (Conjecture \ref{conj2'}); indeed, we have more evidence for its asymptotic validity than for its validity over the entire marked moduli space.}

The organization of this paper is as follows. In Section \ref{sec:cc}, we will review the Cobordism Conjecture (CC) of \cite{mcnamara_cobordism_2019} as a motivation for the expected triviality of quantum gravity field spaces. We will also precisely define the notion of marked moduli spaces and show that the contractibility of these spaces is plausible (as per Conjecture \ref{conj1}). In Section \ref{sec:q>8}, we will introduce and review several mathematical results relating the curvature and geodesics of manifolds with their topological contractibility. It is from this perspective that we will make contact between our new conjectures and the original negative curvature conjecture of \cite{ooguri_geometry_2007}. We will then apply this to show the contractibility of marked moduli spaces with large amounts of supersymmetry ($Q > 8$ supercharges). 

In Section \ref{sec:q=8}, we will review an argument in \cite{liu_hodge_2016} for the contractibility of vector multiplet moduli spaces in theories with $Q = 8$, and we will see that the unique geodesic property enjoyed by the auxiliary Hodge metric (related to its nonpositive curvatures) plays an important role. In Section \ref{sec:geo}, we will reexamine the physical WP geometry of vector multiplet moduli spaces with $Q=8$ supercharges in the asymptotic regime, providing evidence for Conjecture \ref{conj2'}. We will review an explicit example with badly behaved positive curvature in some asymptotic limits, and we will nevertheless show that Conjecture \ref{conj2'} holds in this class of examples. Further strengthened and refined arguments for the uniqueness of shortest paths in the asymptotic regime across \textit{all} theories with $Q=8$ supercharges are presented in Appendix \ref{sec:dev}. 

In Section \ref{sec:cft}, we will turn our attention to the moduli space interior and provide evidence for Conjecture \ref{conj2} in the vicinity of finite-distance curvature singularities at which a decoupled QFT sector appears and the curvature potentially diverges to $+\infty$. In these examples, we will see how the marking on moduli space is crucially required to ensure that Conjecture \ref{conj2} is satisfied, Moreover, we will see why we need piecewise geodesics connected through the singular points in order to reach arbitrary points on the marked moduli space.  Finally, in Section \ref{sec:concl}, we will discuss the implications and interpretations of our conjectures. 

\section{Cobordism and Moduli Space Topology} \label{sec:cc}

In this section, we will review the Cobordism Conjecture (CC) of \cite{mcnamara_cobordism_2019} and discuss its relationship with the geometry and topology of moduli spaces. Roughly speaking, the CC is the idea that there exists a finite-tension domain wall between any two quantum gravity configurations. More precisely, suppose that $\Omega_D^{\m{QG}}$ is the set of all possible quantum gravity backgrounds in $D$ dimensions. In fact, $\Omega_D^{\m{QG}}$ has the natural structure of an abelian group, where the group operation is given by taking the disjoint union of two theories. The CC is then simply the statement that 
\begin{equation}
    \Omega_D^{\m{QG}} = 0. \label{eq:cc}
\end{equation}
This can be concretely understood in geometric backgrounds for quantum gravity in terms of the mathematical notion of cobordism classes. To make the connection, suppose that we are given a $d$-dimensional theory of quantum gravity which we compactify on a $k$-dimensional internal manifold $M_k$, so that $D=d-k$ (in the notation of \cite{mcnamara_cobordism_2019}). The manifold $M_k$ is equipped with various structures, such as orientation or spin structure, to make it compatible with the $d$-dimensional quantum gravity theory. Denoting this structure by $\m{QG}$, the CC therefore reduces to the statement that
\begin{equation}
    \Omega_k^{\m{QG}} = 0, \label{eq:cc2}
\end{equation}
where $\Omega_k^{\m{QG}}$ is now the abelian group of cobordism classes of $k$-manifolds with ``quantum gravity structure'' whose group operation is given by disjoint union.

The CC is a refinement of the classical no-global-symmetries conjecture \cite{banks_symmetries_2011, polchinski_monopoles_2004,misner_classical_1957, harlow_symmetries_2019} in the following sense: The existence of a nontrivial class in $\Omega_k^{\m{QG}}$ implies a $(d-k-1)$-form global symmetry (as per \cite{gaiotto_generalized_2015}). Indeed, one could consider ``gravitational solitons'' formed by gluing a nontrivial class $[M] \in \Omega_k^{\m{QG}}$ to $\B{R}^k$ by excising a small $k$-ball from each of $M$ and $\B{R}^k$ and identifying their respective boundaries. Shrinking the size of $M$, we see that the space looks flat outside a localized region in $(d-k-1)$ dimensions. This $(d-k-1)$-defect therefore carries a global charge labeled by $\Omega_k^{\m{QG}}$, which is forbidden by the no-global-symmetries conjecture. 

In practice, cobordism groups $\Omega_k^{\C{F}}$ have been computed for a vast number of tangential structures $\C{F}$, both in the mathematics \cite{kirby_calculation_1990, giambalvo_langle_1971, anderson_structure_1967} and physics \cite{dierigl_r7-branes_2023, dierigl_iib_2023, basile_global_2024, garcia-etxebarria_dai-freed_2019, freed_reflection_2021, wan_higher_2019, debray_chronicles_2023} literature. These groups rarely vanish, necessitating one of two possible conclusions. If $[M] \in \Omega_k^{\C{F}}$ is a putative nontrivial cobordism class, then one of the following must occur:
\begin{itemize}
    \item The resulting global symmetry is \textit{broken}. In other words, there is a \textit{finite-tension} $(d-k-1)$-defect which trivializes the class $[M] \in \Omega_k^{\C{F}}$. 
    \item The resulting global symmetry is \textit{gauged}. In other words, after including all of the relevant additional structure, the background $[M]$ no longer supports a valid quantum gravity theory.
\end{itemize}
In much of the literature, the first possibility is analyzed in great detail, and physical interpretations have been ascribed to many of the ``cobordism defects'' thus predicted \cite{dierigl_r7-branes_2023, dierigl_iib_2023, basile_global_2024, debray_chronicles_2023, kaidi_non-supersymmetric_2023, montero_cobordism_2021, friedrich_cobordism_2024}. The CC has thus seen remarkable success in both reproducing known objects and predicting new objects in the spectrum of string theory examples. Note, however, that the CC is a purely topological statement -- it tells us nothing about the tension or dynamics of the predicted defects.

\subsection{Triviality of Quantum Gravity Configuration Spaces}

In this section, we will discuss another interpretation of the CC: the triviality of configuration spaces in quantum gravity. As we noted, the CC is a purely topological statement, but this interpretation, as we will see, potentially lends itself to dynamical refinements of the CC. 

First of all, we see that the CC tells us that the configuration space of quantum gravity must be connected. Put another way, if we are given some general space $\C{S}$ parametrizing possible configurations of quantum gravity, then the set of path components of $\C{S}$ should have a single element:
\begin{equation}
    \pi_0(\C{S}) = 0. \label{eq:pi0s}
\end{equation}
In fact, we should expect something even stronger. Compactifying our theory parametrized by $\C{S}$ on some $d$-manifold $\Sigma_l$ amounts to specifying a map $\Sigma_l \to \C{S}$. Thus, letting $[\Sigma_l, \C{S}]$ be the space of maps $\Sigma_l \to \C{S}$, we should have $\pi_0([\Sigma_l, \C{S}]) = 0$. In particular, letting $\Sigma_l = S^d$, we would expect the $d$-th homotopy group of $\C{S}$ to be trivial: 
\begin{equation}
    \pi_l(\C{S}) = 0, \q d \geq 0. \label{eq:pids} 
\end{equation}
If this is true for all $d$, then (excepting pathological mathematical counterexamples) $\C{S}$ must be \textit{contractible}. That is, there is a retraction from $\C{S}$ back to any point $p \in \C{S}$. Contractibility is in some sense the strongest homotopy-theoretic triviality condition, and CC tells us that it would be natural to expect this strong form of triviality to be true of quantum gravity configuration spaces. 

Thus far, our discussion has been entirely topological. However, gravity is not topological, and its configuration spaces are influenced by dynamical quantities, such as the masses of its particles and the tensions of its extended objects. Thus, the characterization of the exact space predicted to be trivial by the CC is not always easy. In particular, the CC tells us of the existence of finite-tension defects that trivialize cobordism classes, but it tells us nothing about the actual tension of these defects. Thus, if we consider the configuration space $\C{S}_\Lambda$ below some cutoff scale $\Lambda$, it could very well have nontrivial topology, but as we take $\Lambda \to M_p$, cells are attached to $\C{S}_\Lambda$ to trivialize the topology.\footnote{From a Morse-theoretic perspective, we can think of an cutoff $\Lambda : \C{S} \to \B{R}_{\geq 0}$ as a Morse function on this configuration space. At its critical points, the topology of the space $f^{-1}[0, \Lambda]$ changes by the attachment of new cells.}

Nevertheless, we might expect that some version of the triviality of configuration spaces holds even at finite energy. The easiest place to look is at when the cutoff $\Lambda$ is taken to \textit{zero}; in this case, $\C{S}_\Lambda$ corresponds to the moduli space of vacua $\C{M}$. Thus, one might naively want to show that $\C{M}$ is contractible: 
\begin{equation}
    \pi_d(\C{M})=0, \q d \geq 0 \label{eq:pikM}.
\end{equation}
However, this statement is false for even the simplest examples. Consider Type IIB string theory in 10 dimensions. There is a single complex modulus: the axiodilaton $\tau = C_0 + i e^{-\phi}$. Here, $C_0$ is the RR 0-form field and $\phi$ is the dilaton, related to the string coupling via $g_s = e^\phi$. Notably, Type IIB enjoys a $\PSL(2,\B{Z})$\footnote{Upon including the $(p, q)$-string, this duality group is lifted to $\SL(2,\B{Z})$. After the inclusion of fermions, $\SL(2, \B{Z})$ is in turn lifted to its double covering, $\m{Mp}(2, \B{Z})$. Including as well the worldsheet parity symmetry, the full duality group is $\GL^+(2, \B{Z})$, the $\m{Pin}^+$ cover of $\GL(2, \B{Z})$. See \cite{debray_chronicles_2023} for details.} duality symmetry acting on $\tau$ via
\begin{equation}
    M = \mt{ a & b \\ c & d } : \tau \longmapsto \frac{a\tau + b}{c\tau + d}. \label{eq:sl2ztau}
\end{equation}
The moduli space of $\tau$ is therefore given by $\Gamma \backslash \B{H}$, where $\Gamma = \PSL(2, \B{Z})$ and $\B{H}$ is the upper half-plane. Now, the topological fundamental group $\pi_1(\C{M})$ is trivial, since $\C{M} \simeq \B{R}^2$ as a topological space, forgetting about automorphisms at the fixed points of $\Gamma$.\footnote{Note that the \textit{orbifold} fundamental group of $\C{M}$ is already nontrivial: $\pi_1^{\m{orb}}(\Gamma\backslash\B{H}) = \Gamma$. Here, the orbifold fundamental group keeps track of loops winding around fixed points of the $\SL(2, \B{Z})$ action. For instance, a loop would need to wind three times around the fixed point of order three at $\t = e^{2\pi i/3}$ to be trivial in the orbifold fundamental group. Each nontrivial class in $\pi_1^{\m{orb}}(\C{M})$ corresponds to translation by an element of $\Gamma$. See \cite{dierigl_swampland_2021} for details.} Moreover, it was argued in \cite{ooguri_geometry_2007} that there is no curve of minimum length in each homotopy class on $\C{M}$. Moreover, it was conjectured that a duality group $\Gamma$ is generated by elements of finite order with fixed points. In this case, any 1-cycle in any orbifold homotopy class can be shrunk down to segments looping around each of the fixed points, which have zero length. 

On the other hand, the space $\Gamma\backslash \B{H}$ does not include its so-called ``compactification point'' at the cusp $+i\infty$, whose inclusion corresponds to the lift from Type IIB to F-theory by the inclusion of D7-branes \cite{dierigl_swampland_2021}. Indeed, a sequence of closed paths in any given homotopy class whose lengths decrease to zero can be formed by contracting a loop around the cusp, and we have $\pi_1(\tilde{\C{M}}) = 0$. We almost immediately run into a problem if we try to extend this to higher dimensions: the higher homotopy groups of ${\C{M}}$ do not vanish! For example, even the second \textit{homology} group is nontrivial:
\begin{equation}
    H_2({\C{M}}, \B{Z}) = \B{Z}.
\end{equation}
Homology is a much coarser invariant than homotopy, and indeed the corresponding homotopy group $\pi_2({\C{M}})$ is also nontrivial. From the perspective of the CC, trivializing these homotopy classes corresponds to finding an end-of-the-universe wall for Type IIB on $\B{P}^1$ (F-theory on elliptically-fibered K3). Such a defect must break all supersymmetry and even though it is expected to exist, it is currently unknown how to construct.\footnote{  To make matters worse, as a topological space, ${\C{M}} \simeq S^2$, which has wildly complicated higher homotopy groups. The precise homotopy theory of ${\C{M}}$ is not exactly that of $S^2$ -- it depends on the underlying moduli stack, whose geometric realization is $\Gamma \backslash \B{H}$. Nevertheless, it will be true in general that the higher homotopy groups of ${\C{M}}$ are very nontrivial.}  These configurations predicted by CC will necessarily take us off the massless moduli space and into regions requiring higher energies. The question is now this: Does there exist any notion of topological triviality even in the massless sector? It is this question that leads us to the notion of marked moduli space.

\subsection{Marked Moduli Spaces} \label{sec:mms}

Despite the apparent topological complexity of $\C{M}$ (and completions thereof), one thing stands out. The universal covering space of $\Gamma\backslash\B{H}$ is just $\B{H}$ itself, which \textit{is} contractible. This suggests that the duality group $\Gamma$ is in some sense responsible for the nontrivial topology of $\C{M}$. 

Consider the following setup. Suppose we are given a moduli space of vacua $\C{M}$ in a theory of quantum gravity. Generically, the duality group will be some Higgsed gauge symmetry, so $\C{M}$ may be viewed as the quotient $\Gamma\backslash\hat{\C{M}}$ of some covering space $\hat{\C{M}}$ by the action of the duality group $\Gamma$. Then, by the arguments in the preceding section, it starts to seem plausible that the space $\hat{\C{M}}$ is contractible in the context of quantum gravity. 

We now introduce the notion of a \textit{marked moduli space}, effectively making precise the notion of what $\hat{\C{M}}$ is relative to $\C{M}$. 

\begin{defn}[Marked Moduli Space]
    For a given theory, its \textit{marked moduli space} $\hat{\C{M}}$ is the geometric space whose points correspond to vacua of the theory along with a choice of basis of observables for the theory.
\end{defn}

Returning to the example of Type IIB string theory, the points of its marked moduli space would label vacua of the theory, along with specifying (for instance) which strings are to be labeled as the F1/D1-strings. Since all of these labels are related by a transformation in $\SL(2,\B{Z})$ (or some extension thereof), the marked moduli space is simply parametrized by the axiodilaton vacuum expectation value (vev) without the subsequent quotient by $\Gamma=\PSL(2,\B{Z})$. Thus, $\hat{\C{M}} = \B{H}$ and is contractible, as expected. We note that the marked moduli space $\hat{\C{M}}$ does not include the singular points (at finite or infinite distance) included in the unmarked moduli space $\C{M}$ (for instance, the lift of the cusp point $+i\infty$ in the Type IIB example is not included in $\hat{\C{M}} = \B{H}$).  

The concept of marked moduli space is understood more broadly in the context of Calabi-Yau (CY) manifolds. Given a CY $n$-fold $X$, a marking is a choice of basis for $H^n(X, \B{Z})/\m{Tors}(H^n(X, \B{Z}))$, the free part  of $H^n(X, \B{Z})$ (here $\m{Tors}(A)$ denotes the torsion components of an abelian group $A$) \cite{liu_global_2016, liu_hodge_2016}. The \textit{Teichmüller space} $\hat{\C{M}}$ of a marked CY $n$-fold $X$ is the space parametrized by complex structure moduli of the marked CY $n$-fold, and is what corresponds to our notion of a marked moduli space.\footnote{In the mathematical literature, a distinction is made between the \textit{Torelli space}, which is defined as the marked moduli space we have considered, and the \textit{Teichmüller space}, which is defined as the universal cover of the moduli space. The Hodge metric completions of these spaces are identical, however, and the distinction between them will be irrelevant for our purposes. Henceforth, we will solely use the term ``Teichmüller space'' to describe the CY marked moduli space. See \cite{liu_global_2016, liu_hodge_2016} for details.}

In the context of CY compactifications of Type IIB, the appropriate metric on $\hat{\C{M}}$, realized as the vector multiplet marked moduli space of the 4d $\C{N}=2$ effective theory, is the Weil-Petersson (WP) metric, defined in this case as the Kähler metric arising from the following potential $K$ \cite{lu_weil-petersson_2005}:
\begin{equation} K = -\log \int_X \Omega \wedge \bar\Omega, \label{eq:wp} \end{equation}
where $\Omega$ is the holomorphic 3-form on $X$. 

Note that the concept of a marked moduli space is a \textit{physical} one which happens to coincide with the \textit{mathematical} notion of Teichmüller space in the case of geometric compactifications. A marked moduli space can be defined for an EFT without any reference to deformations of compact internal geometries. Thus, our Conjectures \ref{conj1}, \ref{conj2'}, and \ref{conj2} are formulated purely with respect to EFTs coupled to quantum gravity. 

We conclude by remarking that the contractibility of marked moduli spaces is \textit{not} simply an analogue of the CC on the space of vacua. Indeed, as we have seen, the unmarked moduli space is not contractible, and we are required to include defects at finite tension above the vacuum in order to trivialize the topology. However, the contractibility of marked moduli space is a statement at the level of vacua itself and knows nothing \textit{a priori} about the finite-tension cobordism defects. We see that our conjectures are made in the same spirit as the CC, but they are in this sense disjoint.

\section{Contractibility of Marked Moduli Spaces: \texorpdfstring{$Q > 8$}{Q>8} Supercharges} \label{sec:q>8}

In this section, we will address the problem of contractibility of marked moduli spaces in cases with high supersymmetry ($Q > 8$ supercharges). In particular, we will discuss the deep relationship between curvature, geodesics, and contractibility, and we will see how negative curvature should in some sense be seen as a feature of contractible geometric spaces. In doing so, we will come closer to Conjecture \ref{conj2}, the geometric refinement of Conjecture \ref{conj1}. 

The relationship between geodesics and contractibility is seen as follows. Let $M$ be a Riemannian manifold, and pick $p \in X$. The \textit{exponential map} (often called the \textit{geodesic map} in the physics literature) $\exp : T_p X \to X$ is defined by taking the unique geodesic $\g_{\hat{v}}$ pointing along the unit vector $\hat{v} = v/\abs{v} \in T_p X$ and traveling a distance $\abs{v}$ to reach a point $q$. 
\begin{equation}
    \exp : v \mapsto q := \gamma_{\hat{v}}(\abs{v}) \label{eq:expmap}
\end{equation}
If the exponential map is a \textit{diffeomorphism} from its domain onto $M$, then it provides a canonical retraction of $M$ onto $p \in M$. The failure of the exponential map to be injective is therefore an indicator of nontrivial topology on $M$. Thus, if one can show that the exponential map on $M$ is a diffeomorphism at \textit{any} point $p$, the contractibility of $M$ follows. 

Note that the condition that $\exp$ is a diffeomorphism is far stronger than simple topological contractibility. For instance, consider the space $U = S^2\backslash \{p\}$, where $p$ is the north pole, endowed with the standard round metric on $S^2$. Then $U \simeq \B{R}^2$ and is clearly contractible. Nevertheless, $\exp$ is not injective for this example -- there are even closed geodesics on $U$, such as the great circle around the equator. 

The condition that the $\exp$ map at every point is injective \textit{with respect to the physical metric} on marked moduli space is precisely our Conjecture \ref{conj2}. It is in this sense that Conjecture \ref{conj2} represents a ``geometrization'' of Conjecture \ref{conj1}. Note that Conjecture \ref{conj2} argues only for the \textit{injectivity} of the exponential map, which does not necessarily imply that $\exp$ is a diffeomorphism, which will not generically be the case if $\mathcal{\hat M}$ has finite-distance curvature singularities. In these cases, the uniqueness of shortest paths replaces the bijectivity of the exponential map.

\subsection{Geodesics and Nonpositive Curvature}
\label{sec:ch}

The global geometry of a manifold, encoded in its geodesics, specifies its local geometry, encoded in its curvature. In some cases, it is possible to go in the other direction --- the local geometry can determine the global geometry. The Cartan-Hadamard (CH) theorem \cite{cartan_geometrie_1925, hadamard_surfaces_1898} is precisely such an example:

\begin{thm}[Cartan-Hadamard]
Let $M$ be a simply connected complete Riemannian manifold. Suppose that all the sectional curvatures of $M$ are everywhere nonpositive. Then $M$ is contractible, and moreover, $\exp : T_p M \to M$ is a diffeomorphism for every $p \in M$.
\end{thm}

The intuition for the CH theorem is as follows. Negative curvature tends to push geodesics apart, while positive curvature tends to draw them together. This is seen by inspecting the geodesic deviation equation: Let $z^i(t)$ be a geodesic on a manifold $M$, and let $X^i(t)$ denote the difference between $z^i(t)$ and a nearby geodesic. Then we have the following evolution equation for $X^i$ \cite{carroll_spacetime_2004}:
\begin{equation} \frac{D^2 X^i}{dt^2} = \tensor{R}{^{i}_{jkl}} \frac{dz^j}{dt} \frac{dz^k}{dt} X^l. \label{eq:geodev} \end{equation}
Here $D^2/dt^2$ is the \textit{directional covariant derivative} along the geodesic path $z^i(t)$ and $\tensor{R}{^i_{jkl}}$ is the Riemann tensor on $M$. If the sectional curvatures $R_{ijij} = -R_{ijji}$ are nonpositive, then the matrix multiplying $X^l$ will have nonnegative eigenvalues, and $X^l(t)$ will diverge as $t \to \infty$. In summary, if the sectional curvatures are nonpositive, we find that infinitesimally nearby geodesics will not re-intersect. 

Suppose that we choose points $p, q$ and that there are two distinct vectors $v, v' \in T_p M$ such that $\exp_p(v) = \exp_p(v') = q$. Then there is a continuous path $v(s)$ such that $v(0) = v$ and $v(1) = v'$ on the sphere of radius 1 in the tangent space $T_p M$. Now, take the exponential map $\exp(v(s))$. For $s$ arbitrarily close to zero, we know that $\exp(v(s)) \neq p$, since the negative curvature pushes nearby geodesics apart. We thus conclude that $\exp(v(s))$ traces out a \textit{nontrivial} closed loop containing $q$. Since $M$ is simply connected, this loop can be continuously deformed to $q$ itself. We now find a contradiction --- in the preimage of the exponential map (which is nonempty since $M$ is complete) there is no corresponding continuous deformation of the path $v(s)$ to a single point that preserves both endpoints $v(0) = v$ and $v(1) = v'$. 

Note the importance of the simply-connectedness of $M$ in the above argument --- if $M$ had a nontrivial fundamental group, then one could easily construct separated re-intersecting geodesics labeled $v, v' \in T_p M$ from a point $p$. The corresponding loop traced out by $\exp_p(v(s))$ would then have a nontrivial homotopy class, in which case it cannot be deformed to the point $q$. In effect, this tells us that manifolds with negative curvature can have a nontrivial fundamental group, but their higher homotopy groups must be trivial. This also makes intuitive sense --- the higher spheres $S^n$ for $n \geq 2$ must have regions of positive curvature (as can be seen for $S^2$ by, for instance, the Gauss-Bonnet theorem), so it makes sense that we cannot geometrically ``wrap'' them around a globally negatively curved space $M$. On the other hand, $S^1$ is flat, so there is no problem considering nontrivial maps $S^1 \to M$. 

The CH theorem provides an important connection between \textit{global} topology and geometry (contractibility and the unique geodesic property) and \textit{local} geometry (nonpositive curvature). It is in this way that it links our Conjectures \ref{conj1} and \ref{conj2'} with the asymptotic negative curvature conjecture in \cite{ooguri_geometry_2007}. The authors of \cite{ooguri_geometry_2007} conjectured that the scalar curvature in asymptotic regions at infinite distance in (unmarked) moduli spaces is always nonpositive. Although this exact statement was seen to be false in \cite{trenner_asymptotic_2010} (with the relevant counterexamples later elaborated on in \cite{marchesano_moduli_2023}), we see that the spirit of the old conjecture is preserved in Conjectures \ref{conj1} and \ref{conj2'} via the CH theorem. In effect, we believe that the asymptotic negative curvature observed in numerous examples should be viewed as a \textit{consequence} of a larger principle: the uniqueness of geodesics on marked moduli space.

Using the CH theorem, it is not difficult to argue for the contractibility of moduli spaces in theories with more than $8$ supercharges. In the remainder of this section, we will review these examples with high supersymmetry and show that their moduli spaces indeed have negative curvature in their canonical metric.

\subsection{Contractibility with \texorpdfstring{$Q > 8$}{q>82}}

In this section, we will argue for the contractibility of moduli spaces with $Q > 8$ supercharges. To make matters simple, we will discuss the details only for $Q = 32$ and $Q = 16$ supercharges, with the other cases being entirely analogous \cite{cecotti_swampland_2021}.

Let us first address the case of maximal supersymmetry with $Q=32$ supercharges. Such examples arise via simple toroidal compactification of M-theory. Compactifying on $T^n$, the appropriate moduli space of the scalars is given as follows \cite{cremmer_n8_1978, cremmer_so8_1979, julia_group_1981, de_wit_local_1985}:

\begin{equation} \C{M} = E_{n(n)}(\B{Z}) \backslash E_{n(n)}/ K_n, \end{equation}
where $E_{n(n)}$ is the split real form of $E_n$, $K_n$ is its maximal compact subgroup, and $E_{n(n)}(\B{Z})$ is the discrete U-duality group. (See \cite{fre_general_2006} for a review of U-duality and moduli spaces in maximal supergravity). We claim that the associated marked moduli space is then
\begin{equation} \hat{\C{M}} = E_{n(n)} / K_n . \end{equation}
This is a connected, simply-connected irreducible Riemannian symmetric space of noncompact type. Its metric descends from the Killing form on $E_{n(n)}$, and it is a classical result that all such spaces have constant nonpositive sectional curvature \cite{eberlein_structure_1997}.

The case with $Q=16$ supercharges is equally easy. Such examples are known to descend from toroidal compactifications of Heterotic string theory from 10 dimensions. The associated moduli spaces are the Narain moduli spaces, which take the following general form \cite{narain_new_1986, narain_note_1987}:
\begin{equation} \C{M} = \SO(r, n; \B{Z})\backslash \SO(r, n; \B{R})/ (\SO(r) \times \SO(n) ). \label{eq:narain} \end{equation}
Here $r$ is the rank of the gauge group and $n$ is the number of compact dimensions. The exact values of $r, n$ depend on the theory in question, but all moduli spaces with 16 supercharges take the above form. We claim that associated marked moduli spaces are
\begin{equation} \hat{\C{M}} = \SO(r, n; \B{R})/ (\SO(r) \times \SO(n) ),  \end{equation}
which is again a connected, simply connected Riemannian symmetric space of noncompact type with nonpositive sectional curvatures. The negative curvature of these moduli spaces has been observed long ago, and was one of the original motivations for the asymptotic negative curvature conjecture in \cite{ooguri_geometry_2007}.

We have intuitively deduced that the marked moduli spaces are just coverings of the unmarked moduli spaces, without taking a quotient by the $U$-duality group. We will illustrate why this is the case for these examples. Consider the $Q = 16$ example. The vector-multiplet moduli are then the Wilson lines wrapping cycles in the compact geometry $T^d$. A choice of marking in this example is given by a basis for the charge lattice of the gauge group. But the automorphism group for the charge lattice is given precisely by $\SO(r, n; \B{Z})$. The marked moduli space is therefore a covering space of $\C{M}$, the deck transformation group of which is $\SO(r, n; \B{Z})$. In other words, it is precisely the space $\SO(r,n ; \B{R})/ (\SO(r) \times \SO(n) )$ before taking a quotient by $\SO(r,n; \B{Z})$, as claimed. We have only illustrated this for our chosen examples, but all of the other cases for $Q > 8$ supercharges are analogous \cite{cecotti_swampland_2021}.

\section{Contractibility of Marked Moduli Spaces: \texorpdfstring{$Q = 8$}{q-8} Supercharges} \label{sec:q=8}

The contractibility of marked moduli spaces with high supersymmetry was straightforward enough to establish. Indeed, the moduli spaces with high supersymmetry are geometrically simple --- they are symmetric spaces of noncompact type. However, the picture becomes more complicated as soon as one reduces to $Q = 8$ supercharges. In what follows, we will focus on the vector multiplet moduli spaces in theories with $Q=8$ supercharges. These moduli spaces are easier to study than the hypermultiplet moduli spaces, which receive large quantum corrections. 

We will consider the case of a 4d $\C{N}=2$ theory obtained by compactification of Type IIB on a CY3. By mirror symmetry, the large-complex-structure limit of this moduli space is equal to the asymptotic large-volume regime of the 5d $\C{N}=1$ vector multiplet moduli space of M-theory on the mirror CY3. In the 5d $\C{N}=1$ case of M-theory on a CY3 $X$, the vector multiplet moduli space is parametrized by the Kähler moduli of $X$ spanning the Kähler cone, which is manifestly contractible. In the 4d $\C{N}=2$ case of Type IIB on the mirror $X^\vee$, the vector multiplet moduli space is parametrized by the complex structure moduli of $X^\vee$. The marked moduli space then corresponds to the Teichmüller space $\hat{\C{M}}$ of $X^\vee$, and the metric arises from the EFT scalar kinetic terms is the WP metric, defined in Eq. \ref{eq:wp}.
In this section, we will argue for the contractibility of $\hat{\C{M}}$, making use of an auxiliary metric. 

The WP metric is not the only metric one can place on CY3 complex structure moduli space $\C{M}$. It is possible to write down a second canonically-defined metric, the \textit{Hodge metric}, on $\C{M}$ (and therefore the Teichmüller space $\hat{\C{M}}$). The Hodge metric has nonpositive curvature properties and can be used to show the topological contractibility of the $\hat{\C{M}}$. In this section, we will review the results in \cite{liu_hodge_2016} and previous work to establish the contractibility of Teichmüller space using the properties of its Hodge geometry. 

The Hodge metric is defined on $\hat{\C{M}}$ via the calculation of periods. The period map $\psi : \hat{\C{M}} \to D$ sends each point $p \in \C{M}$ to the corresponding point in the space of periods $D$ by computing the period integrals on the corresponding CY3. A point in the space $D$ is calculated by computing the period integrals of the CY3 about a basis of its 3-cycles. Under the polarization induced by the complex structure defined on the CY3, the canonical intersection pairing between 3-cycles becomes a symplectic bilinear form. Thus, the period space $D$ is homogeneous with respect to a group action preserving this symplectic intersection pairing between 3-cycles in the CY3.

For $h^{2, 1} = n$, $D$ is therefore a quotient $G_{\B{R}}/V = \Sp(2n+2, \B{R})/(\U(n)\times \U(1))$ where $G_{\B{R}}=\Sp(2n+2, \B{R})$ is a noncompact real Lie group and $V = B \cap G_{\B{R}} = \U(n) \times \U(1)$ is the intersection of $G_{\B{R}}$ with a Borel subgroup $B$ of the complexified group $G = \Sp(2n+2, \B{C})$ \cite{cecotti_swampland_2021}. Thus, $D$ is a homogeneous space of noncompact type with a canonical invariant metric induced by the Killing form on $G_{\B{R}}$. In particular, $D$ is simply connected \cite{griffiths_locally_1969}. The pullback of this metric along the period map $\psi$ therefore furnishes a metric on $\hat{\C{M}}$, the Hodge metric, which we denote by $k_{i\bar j}$. 

The curvature of $k_{i\bar j}$ is much better-behaved than that of the WP metric $g_{i\bar j}$ in the following sense. It is computed in \cite{lu_geometry_2005, lu_hodge_2005} that its holomorphic bisectional curvatures are nonpositive. In fact, the Ricci curvature $R^k_{i\bar j}$ is nonpositive and bounded \textit{away} from zero: 
\begin{equation} R^{k}_{i\bar j} \leq -\frac{1}{(\sqrt{n + 1} + 1)^2} k_{i\bar j}. \end{equation}
The reader is directed to \cite{lu_hodge_2005} for detailed calculations. Note that the nonpositivity of the holomorphic bisectional curvature, while related to the unique geodesic property \cite{lu_hodge_2005, milnor_curvatures_1976,azencott_homogeneous_1976,noauthor_topics_1984} does not imply the nonpositivity of Riemannian sectional curvature as required by the CH theorem.\footnote{In fact, there are examples of simply connected Kähler manifolds with nonpositive holomorphic bisectional curvatures which are not contractible \cite{mohsen_construction_2019}.}  Thus, one must do more work to show the contractibility of $\hat{\C{M}}$.

In \cite{liu_hodge_2016, liu_curvatures_2016}, the authors show that the marked moduli space $\hat{\C{M}}$ with respect to the Hodge metric is a connected and simply-connected complete complex affine manifold, which is in particular contractible and diffeomorphic to Euclidean space.\footnote{To be precise, it is shown that the \textit{completion} $\hat{\C{M}}^H$ of the marked moduli space $\hat{\C{M}}$ with respect to the Hodge metric is diffeomorphic to Euclidean space. In general, it is the space $\hat{\C{M}}^H$ that should be considered as the ``true'' marked moduli space. Now, curvature singularities related to the incompleteness of the physical WP metric are pushed to infinite distance in the Hodge metric, so the ``completion points'' with respect to the Hodge metric lie at infinite distance, and effectively $\hat{\C{M}} = \hat{\C{M}}^H$ in the examples we have considered.} We will sketch the salient features of their argument below; the reader is encouraged to consult the original reference \cite{liu_hodge_2016} (in particular, Theorem 3.10, Lemma 3.11 and the preceding discussion) for a fully detailed and rigorous proof. 

Let $\psi : \hat{\C{M}} \to D$ be the period map as defined previously. As part of the proof in \cite{liu_hodge_2016}, the structure of $D$ as a homogeneous space is used to show that the image of the tangent space at an arbitrary $p \in \hat{\C{M}}$ under the period map $\psi$ is an abelian Lie algebra $\f{a}_0 \subset T_{\psi(p)} D$:
\begin{equation} \f{a}_0 = \psi_*(T_p \hat{\C{M}}) , \end{equation}
where $\psi_*$ is the tangent map of $\psi$. Since $D$ is a quotient of $G_{\B{R}} = \Sp(2n+2, \B{R})$, the Lie algebra $\f{a}_0$ arises as a quotient of the split real form $\f{sp}(2n+2)_{\B{R}}$. 

The map $\exp : \f{a}_0 \to \exp(\f{a}_0)$ is a diffeomorphism for $\f{a}_0$ an Abelian Lie algebra and $\exp(\f{a}_0)$ the corresponding simply-connected Lie group. Since $D$ is simply-connected, we find that $\exp(\f{a}_0) \into D$ is an injection. The key result (Theorem 3.10 in \cite{liu_hodge_2016}) is that $\psi$ is a biholomorphic map onto its image. From a physical perspective, this is simply the statement that the period integrals ``faithfully'' capture the whole marked CY geometry. It is then seen that there is a diffeomorphism
\begin{equation} \psi(\hat{\C{M}}) \simeq \exp(\f{a}_0) \simeq \f{a}_0. \label{eq:period_map_biholomorphism}\end{equation}
In particular, $\hat{\C{M}}$ is diffeomorphic to Euclidean space and is therefore contractible. We can say much more, however. Recalling that the Hodge metric is the pullback of the metric on $D$ to $\hat{\C{M}}$, Eq. \eqref{eq:period_map_biholomorphism} shows that the exponential map at any $p \in \hat{\C{M}}$ is a diffeomorphism from its domain onto $\hat{\C{M}}$. In particular, the unique geodesic property holds for the Hodge geometry of $\hat{\C{M}}$! An explicit proof of this is provided in further detail in Lemma 3.11 of \cite{liu_hodge_2016}. 

On the other hand, the global Hodge geometry of $\hat{\C{M}}$ is also related to the nonpositivity of Hodge curvatures. It is discussed in \cite{griffiths_locally_1969} that the embedded space $\psi(\hat{\C{M}}) \subset D$ has some of the properties of a Hermitian symmetric space of noncompact type with nonpositive holomorphic bisectional curvatures (note that the entirety of $D$ itself is not necessarily a Hermitian symmetric space). Thus, the most notable \textit{local} geometric feature of the Hodge metric is seen to directly relate to the uniqueness of Hodge geodesics on $\hat{\C{M}}$. We emphasize also that the unique geodesic property of the Hodge metric emerges immediately from the proof of the contractibility of Teichmüller space---Curvature, geodesics, and global topology are closely intertwined. It is therefore natural to expect that the some refinement of the unique geodesic property is indeed the proper geometric strengthening of the topological contractibility condition. 

 The Hodge metric does not \textit{a priori} admit a natural physical interpretation. In the CY3 case, the Hodge metric has a simple relation to the WP metric, first described in \cite{lu_hodge_2005}: 
\begin{equation} k_{i\barj} = (n+3) g_{i\barj} + R_{i\barj},  \end{equation}
where $R_{i\barj}$ is the Ricci tensor of the WP metric $g_{i\barj}$ and where $n = h^{2, 1}$ as before. Various interpretations for this expression in terms of the WP metric were proposed by Cecotti \cite{cecotti_moduli_2020, cecotti_swampland_2021, cecotti_special_2020}.
Nevertheless, the Hodge metric remains unnatural from the EFT perspective. A schematic reason for this is as follows. Points at which $R_{i\barj}$ diverges to $+\infty$ lie at finite geodesic distance in the WP metric corresponding to the shrinking of a 4-cycle on the corresponding CY to zero size, and these points correspond to the emergence of a QFT sector decoupled from gravity \cite{marchesano_moduli_2023}. However, in the geometry of the Hodge metric, these points are pushed off to infinite distance. Since these cycles can clearly shrink to zero with finite action in the effective theory, it appears that the viability of any EFT interpretation of the Hodge metric is limited. 

Thus, the goal remains to understand the WP geometry of the Teichmüller space, as this is what the EFT sees. The space $\hat{\C{M}}$ is not complete for the WP metric due to the generic presence of finite-distance curvature singularities, so the exponential map will generally not be defined over every point in $T_p \hat{\C{M}}$ for $p \in \hat{\C{M}}$. Nevertheless, one might ask if the exponential map is \textit{injective} where it is defined --- in other words, one might ask if the exponential map is a diffeomorphism from its domain, a star-shaped region in $T_p \hat{\C{M}}$, into $\hat{\C{M}}$. Since the unique geodesic property for the Hodge metric emerges in the proof of contractibility of Teichmüller space, we wish to see if a version of this geometric condition also holds for the physical WP metric. This is precisely the question posed by our Conjecture \ref{conj2}: we conjecture that the exponential map with respect to the physical metric remains injective, so that there is at most one geodesic between any pair of points. Moreover, as we will see in Sec. \ref{sec:cft}, we also conjecture that $\hat{\C{M}}$ satisfies the \textit{unique shortest path property} (a refinement of the unique geodesic property to spaces with finite-distance singularities): It is possible to find a unique piecewise-geodesic shortest path between \textit{any} two points in $\hat{\C{M}}$, which may connect through finite-distance curvature singularities. It is the objective of Section \ref{sec:geo} to provide evidence for this at least in the asymptotic regime (our Conjecture \ref{conj2'}), while we will provide evidence for the stronger Conjecture \ref{conj2} in the moduli space interior in Sec. \ref{sec:cft}.  

To conclude, we emphasize that the injectivity of the exponential map, along with the existence of unique piecewise-geodesic shortest paths between any two points, in some sense represents a ``geometrization'' of the topological statement of contractibility. We have already seen in this section that CY3 moduli spaces are contractible in the topological sense, and we wish to show that the associated \textit{geometric} properties continue to hold for the physical metric. 

\section{Unique Geodesics in the Asymptotic Regime} \label{sec:geo}

In this section, we will provide evidence in the asymptotic regime towards our Conjecture \ref{conj2'}.\footnote{The 4d and 5d examples here also provides evidence for \ref{conj2}, as the positive divergent curvature in the asymptotic region survives in the interior of moduli space as well. Indeed, the 5d version of this example is at finite distance in moduli space of vector multiplets, and it can furnish an example of positive divergence at infinite distance only if we consider the overall volume modulus, which is a hypermultiplet.} In Section \ref{sec:positivewp}, we review an example of M-theory compactified on CY3 from \cite{trenner_asymptotic_2010} in which the curvature is not only positive but also \textit{divergent} and that this divergence survives in certain asymptotic regions. However, we will see in \ref{sec:unique} that the unique geodesic property, and therefore Conjecture \ref{conj2'}, is satisfied for this example in the vicinity of the divergent positive curvature. Further discussion and evidence for the unique geodesic property on asymptotic moduli spaces is given in Appendices \ref{sec:ex} and \ref{sec:dev}.  

\subsection{Moduli Spaces with Positive Asymptotic Curvature} \label{sec:positivewp}
 
Consider M-theory on CY3, resulting in an effective 5d $\C{N}=1$ supergravity theory in the low-energy limit. Then the vector multiplet moduli space of the effective theory is the so-called Kähler cone, parametrized by the Kähler moduli of the CY3. (Note that in the 5d theory, the overall volume is part of a hypermultiplet. Under compactificaiton on a circle to 4d, the full Kähler moduli would be part of the vector multiplet. As such, we will be cavalier as to whether we are discussing the 4d or the 5d theory.) We wish to consider the asymptotic regime of the moduli space in which the volume of the CY3 is taken to be large. In \cite{trenner_asymptotic_2010}, several examples were constructed in which there exist limits in the asymptotic geometry with divergent positive curvature. We will review these examples in this section. 

Write the Kähler form $J$ in a basis $\omega_1, \cdots, \w_n$ of closed $(1, 1)$-forms as 
\begin{equation}
    J = y^i \w_i,
\end{equation}
The Kähler moduli $y^i$ then parametrize the 5d $\C{N}=1$ vector multiplet moduli space (up to the overall volume). The local moduli space geometry is specified by a Kähler metric with Kähler potential $-\log C$, where the prepotential $C$ is given in the asymptotic large-volume limit by 
\begin{equation} C = C_{ijk} y^i y^j y^k + \cdots. \label{eq:lvpre} \end{equation}
Here, the $C_{ijk}$ are the totally symmetric triple intersection numbers of the CY3, given by 
\begin{equation} C_{ijk} = \int_X \w_i \wedge \w_j \wedge \w_k. \end{equation}
In terms of the $C_{ijk}$, the Kähler metric is given by 
\begin{equation} g_{ij} = \frac{9}{4} \frac{C_i C_j}{C^2} - \frac{3}{2} \frac{C_{ij}}{C}, \end{equation}
where we have defined
\begin{equation} C_{ij} := C_{ijk} y^k, \q C_i  = C_{ijk} y^j y^k. \label{eq:cab} \end{equation}
For future reference, we also quote a formula for the inverse metric: 
\begin{equation}
     g^{ij} = 2 y^i y^j - \frac{2}{3} C C^{i j},  \label{eq:inversemetric}
\end{equation}
where $C^{ij}$ is the inverse to $C_{ij}$. Note that we are using the \textit{real} indices $i, j$ since we are working in the asymptotic regime (in the 4d sense) where only the saxionic coordinates $y^i$ are relevant. Throughout the rest of this section and what follows, we will calculate purely in terms of these real variables, even though there is an underlying complex metric on the entirety of the Kähler moduli space in the 4d context. The coefficients $C_{ijk}$ can be chosen to be nonnegative in an appropriate basis of divisors (the so-called Nef basis), but the above formula holds for \textit{any} basis in the region that the metric $g_{i\barj}$ is positive-definite. 

All of the examples considered in \cite{trenner_asymptotic_2010} have $h^{1, 1} = 3$. A simple example is given by a hypersurface $V_{16} \subset \B{P}_{1, 1, 1, 5, 8}$, for which the Kähler cone geometry is specified by the following cubic prepotential in the large-volume limit: 
\begin{equation} C(y_1, y_2, y_3) = 50 y_1^3 + 30 y_1^2 y_2 + 6 y_1 y_2^2 + 240 y_1^2 y_3 + 96 y_1 y_2 y_3 + 9 y_2^2 y_3 + 384 y_1 y_3^2 + 75 y_2 y_3^2 + 203 y_3^3. \label{eq:pre1} \end{equation}
We note that this prepotential can be written in the following form: 
\begin{equation} C = \frac{2}{5} (5y_1 + y_2 + 8y_3)^3 - \frac{1}{15} (y_2 + 3y_3)^3 - \frac{1}{3} y_2^3. \end{equation}
Thus, by making a linear change of coordinates, it suffices to consider
\begin{equation} C = z_1^3 - z_2^3 - z_3^3. \label{eq:sec5example}\end{equation}
The metric associated to this prepotential is positive-definite in the region given by 
\begin{equation} z_1, z_2, z_3 > 0, \q z_1^3 > z_1^3 + z_2^3. \end{equation}
It is straightforward to compute the metric determinant and the Ricci scalar: 
\begin{equation} \det g_{ij} = \frac{27 z_1 z_2 z_3}{16 (z_1^3 - z_2^3 - z_3^3)^3}, \end{equation}
\begin{equation} R = -32 + \frac{2}{3} \left[ z_1^3 \left(\frac{1}{z_2^3}  + \frac{1}{z_3^3} \right) + \frac{z_2^3 + z_3^3}{z_1^3} - \frac{z_2^6 + z_3^6}{z_2^3 z_3^3}  \right].  \end{equation}
As we take the infinite distance limit $z_1 \gg z_2, z_3$, the scalar curvature indeed blows up to $+\infty$. Correspondingly, the faces $z_2 = 0$ and $z_3 = 0$ have curvature singularities with positive, divergent curvature. 

It can be checked that the Kähler cone (given by the conditions $y_i \geq 0$) lands within the region of allowed $z_i$ for which the metric is positive-definite. Since $y_2 = z_2$, we see that the face $z_2 = 0$ includes a boundary the Kähler cone geometry. This face turns out to be the one responsible for the curvature divergence. It therefore suffices to study geodesics in the full region described by $z_1, z_2, z_3 > 0$, $z_1^3 > z_1^3 + z_2^3$.

Note that other examples of so-called curvature divergences on dimension-two loci also arise from linear transformations of the prepotential $f = z_1^3 - z_2^3 - z_3^3$ for which one of the faces of the Kähler cone lies on either the face $z_2 = 0$ or $z_3 = 0$. For example, a similar geometry given by a hypersurface $V_{12} \subset \B{P}(1,1,1,3,6)$ has the following prepotential:
\begin{equation} C(y_1, y_2, y_3) = 18 y_1^3 + 18 y_1^2 y_2 + 54 y_1^2 y_3 + 6 y_1 y_2^2 +  36 y_1 y_2 y_3 +  54 y_1 y_3^2 + 3 y_2^2 y_3 + 9 y_2 y_3^2 + 9 y_3^3 . \end{equation}
This can also be put in the form $C = z_1^3 - z_2^3 - z_3^3$ by an appropriate coordinate transformation.

Although we will only consider the cases above with $C = z_1^3 - z_2^3 - z_3^3$ in the remainder of this section, these do not account for all three-moduli examples with divergent positive curvature. Nevertheless, we will provide a general argument in Appendix \ref{sec:dev} that nearby geodesics do not re-intersect near the curvature singularity for \textit{any} geometry specified by $C_{ijk}$ with an \textit{arbitrary} number of moduli. 

The positive curvature in all of these examples is explained as follows. It was argued in \cite{trenner_asymptotic_2010} that for a prepotential $C$ with $h^{1, 1} = n$, we have
\begin{equation} \det g_{ij} = A \frac{\det C_{ij}}{C^n}, \end{equation}
where $A$ is a constant depending only on $n$ and $C_{ij}$ is defined in Eq. \eqref{eq:cab}. The offending divergence of the scalar curvature $R$ can be seen from the formula in \cite{viaclovsky_kahler_2018} for the curvature of a Kähler metric: 
\begin{equation} R = -g^{ij} \p_i \p_{j} \log \det g_{ij} \simeq -g^{ij} \p_i\p_{j} \log \det C_{ij} + \cdots . \end{equation}
Indeed, as one approaches the locus where $h:= \det C_{ij} = 0$ but $C$ remains finite, the curvature diverges to $+\infty$. This locus signals the vanishing of an eigenvalue of $g_{i\barj}$. It therefore lies at finite geodesic distance and is a ``smoking gun'' for the appearance of new physics. Physically speaking, it corresponds to certain 4-cycles shrinking to zero size at finite distance in field space. Note that the locus $h = 0$ also represents a finite-distance \textit{boundary} for the WP geometry, since the metric has a zero eigenvalue here. Thus, we see that the positive curvature is directly tied to the failure of the WP metric to be complete. 

For many examples, the locus $h = 0$ only intersects the Kähler cone at the zero-volume point, so there is no positive curvature divergence. However, the cases described in \cite{trenner_asymptotic_2010} and then considered in \cite{marchesano_moduli_2023} all arise from examples in which $h$ vanishes at a locus of positive dimension at the boundary of the Kähler cone. In the example of Eq. \eqref{eq:pre1}, the locus $h = 0$ has codimension one (dimension two). 

\subsection{Unique Geodesics on Examples with Positive Asymptotic Curvature} \label{sec:unique}

The positive WP curvature is related to the lack of geodesic completeness of $\hat{\C{M}}$ and not \textit{a priori} to any ``spherical'' behavior, so intuitively it seems like the geodesics emanating from a point should throw themselves towards the finite-distance boundary before they have a chance to intersect again. Thus, we expect that the exponential map will not be defined everywhere on $T_p \hat{\C{M}}$ for a given $p \in \hat{\C{M}}$, but it still seems plausible that it should be \textit{injective} from its domain, a star-shaped region, into $\hat{\C{M}}$. 

In this section, we will argue that this is indeed the case for the three-moduli example presented in Eq. \eqref{eq:sec5example} of Section \ref{sec:positivewp}. We will illustrate the general argument in this section; a rigorous argument for unique geodesic property \textit{everywhere} in the Kähler cone for this example is given in Appendix \ref{sec:ex}. In addition, an entirely analogous and more general version of this argument applicable to divergent positive curvature regions for any special Kähler geometry specified by the coefficients $C_{ijk}$ is presented in Appendix \ref{sec:dev}.

To show the injectivity of the exponential map, we argue as follows. Note that $\hat{\C{M}}$ is a simply-connected (indeed, contractible) space. Choose an arbitrary point $p \in \hat{\C{M}}$, and suppose that two distinct geodesics $\g_v, \g_{v'}$ with tangent vectors $v, v'$ emanating from $p$ re-intersect at $q$. Then the exponential map $\exp : T_p \hat{\C{M}} \to \hat{\C{M}}$ is not injective beyond a certain radius in its domain. Thus, a path $v(s)$ in $T_p \hat{\C{M}}$ from $v$ to $v'$ with $v(0) = v$ and $v(1) = v'$ must get mapped to a closed loop $\exp(v(s))$ containing $q$ in $\hat{\C{M}}$, which is contractible since $\hat{\C{M}}$ is simply connected. In the argument for the CH theorem in Section \ref{sec:ch}, it was shown that such a loop could not be nontrivial, since contracting it to zero would yield a contradiction. 

Thus, the only possible outcome is that the entire path is mapped to the single point $q$, so there exists a geodesic arbitrarily close to $\g_v$ that also intersects $q$. In the proof of the CH theorem, we argued using the geodesic deviation equation that such nearby geodesics cannot exist if the sectional curvatures are nonpositive. It therefore suffices to show the same for our present example: that \textit{nearby} geodesics do not re-intersect. For this, we may use the geodesic deviation equation.

Recall the geodesic deviation equation from Eq. \eqref{eq:geodev}: 
\begin{equation} \frac{D^2 X^i}{dt^2} = \tensor{R}{^{i}_{jkl}} \frac{dz^j}{dt} \frac{dz^k}{dt} X^l. \label{eq:geodev1} \end{equation}
where $z^i(t)$ is any geodesic and $X^i(t)$ represents an infinitesimal deviation to nearby geodesics about the geodesic $z^k(t)$. Expanding this expression out, we find that
\begin{equation} \frac{d^2 X^i}{dt^2} + 2\Gamma^i_{jk} \frac{dz^j}{dt} \frac{dX^k}{dt} + \p_{l} \Gamma^i_{jk} \frac{dz^j}{dt} \frac{dz^k}{dt} X^{l} = 0 . \label{eq:geodev2} \end{equation}
It is this equation that we wish to solve for $X^i(t)$ along a geodesic path $z^k(t)$. 

Let us now return to the example from Section \ref{sec:positivewp}. Recall that the prepotential is given by 
\begin{equation} C = z_1^3 - z_2^3 - z_3^3. \end{equation}
In Appendix \ref{sec:prel}, the geodesic equation on an arbitrary Kähler cone geometry in the asymptotic regime is derived. We simply quote the results for our example Eq. \eqref{eq:sec5example}:
\begin{equation} 2 z_i ( z''_i -\a z'_i) + {z'_i}^2 = \l z_i^2, \q i = 1, 2, 3 \end{equation}
along with the constraint
\begin{equation} z_1^3 - z_2^3 - z_3^3 = C_0 e^{\a t}.  \end{equation}
Here, $\l$ is a nonnegative constant. This is solved analytically to yield
\begin{equation} z_i(t) = e^{\a t/3}(A_i e^{\b t} + B_i e^{-\b t})^{2/3}. \label{eq:generalsolution} \end{equation}
where $\a, \b$ are arbitrary real numbers and the $A_i, B_i$ are real numbers satisfying the constraints
\begin{equation} A_1^2 = A_2^2 + A_3^2, \q B_1^2 = B_2^2 + B_3^2 . \end{equation}
The factor $e^{\a t/3}$ is simply an overall scaling present in all geodesics, and does not affect whether nearby geodesics will re-intersect in the geodesic deviation equation. It therefore suffices to consider the case when $\a = 0$. Physically, this corresponds to holding the overall volume $C = z_1^3 - z_2^3 - z_3^3$ constant, which in the 5d case corresponds to moving only in the 2-dimensional vector multiplet moduli space.
Note that this locus is approached in \textit{finite} time from any point in the interior of the geometry, precisely as we argued earlier. 

Let us now study infinitesimal deviations about a given geodesic. We will first consider a geodesic at fixed volume approaching the locus $z_2 = 0$. We show that infinitesimally nearby geodesics do not re-intersect arbitrarily close to $z_2 = 0$. This is in some sense the ``worst case'' regime, as this region has the worst divergent positive curvature. Thus, if the geodesics do not re-intersect here, it is unlikely that they would do so elsewhere. 

\begin{figure}
    \centering
    \includegraphics[scale=0.6]{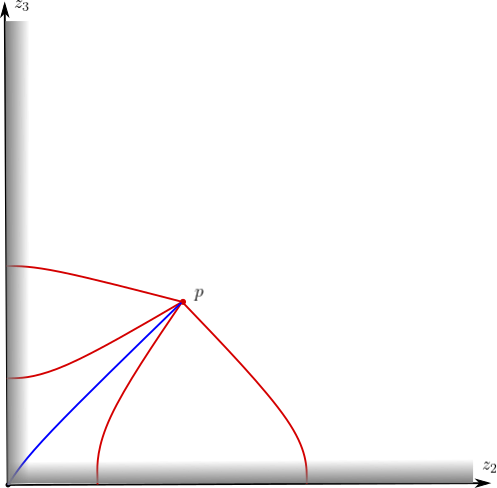}
    \caption{A schematic illustration of various geodesics which approach the singular loci at $z_2 = 0$ and $z_3 = 0$. The red geodesics approach either $z_2 = 0$ or $z_3 = 0$ while leaving the other nonzero; the blue geodesic approaches the origin $z_2 = z_3 = 0$. The gray regions are where the curvature is positive, and we may naively expect nearby geodesics to re-intersect.}
    \label{fig:example-geodesics}
\end{figure}

To this end, fix a point $p = (a, 0, b)$ with $a, b \neq 0$ on the locus $z_2 = 0$, and consider an arbitrary geodesic arriving at this point. Since such a geodesic reaches $p$ in finite time from the interior of the moduli space, let us suppose without loss of generality that $p$ is reached at $t = 0$. Then, in the vicinity of $p$, we will have
\begin{equation} z_2(t) \propto t^{2/3}, \end{equation}
while each of $z_1(t)$ and $z_3(t)$ approach $p$ \textit{linearly} with $t$. Thus, for very small $t$, $z_2 \gg z_1, z_3$, and we may assume that the geodesic is approaching in the direction normal to the plane $z_2 = 0$ and that $z_2(t) \propto t^{2/3}$ as $t \to 0$:
\begin{equation} z^i(t) \simeq (a, bt^{2/3}, c), \q \frac{dz^i}{dt} \simeq \left(0, \frac{2}{3} bt^{-1/3}, 0 \right)  \label{eq:redgeosolution} \end{equation}
A schematic depiction of several different geodesics approaching the $z_2 = 0$ and $z_3 = 0$ planes are shown by the red geodesics in Fig. \ref{fig:example-geodesics}. 

We now wish to study the geodesic deviation equation in the limit as $t \to 0$. In this limit, we are approaching the plane where the scalar curvature diverges to $+\infty$, so it seems \textit{a priori} likely that the unique geodesic condition might be violated here if anywhere at all. Nevertheless, we will show that the geodesic behavior is in some sense completely regular! 

To see this, we extract the most singular terms in $t$ as $t \to 0$ in the coefficients of the geodesic deviation equation. Note that an eigenvalue of $g_{ij}$ is going to zero; correspondingly, an eigenvalue of $C_{ij}$ is going to zero. Thus, the matrix $C^{ij}$ is becoming singular in the limit. 
Substituting this in to the geodesic deviation equation \eqref{eq:geodev2}, we find that
\begin{equation} \frac{d^2 X^i}{dt^2} + \tensor{M}{^i_k}(t) \frac{dX^k}{dt} + \tensor{N}{^i_l}(t) X^l = 0\end{equation}
where
\begin{equation} \tensor{M}{^i_k}(t) = \frac{2}{3t} \mt{ 0  & & \\ & 1 & \\ & & 0 }, \q \tensor{N}{^i_l}(t) = - \frac{2}{9t^2} \mt{ 0  & & \\ & 1 & \\ & & 0 }.  \end{equation}
In particular, the geodesic deviation along directions \textit{orthogonal} to the geodesic path $dz^i/dt$ are nonsingular as $t \to 0$. We also find that the coefficients of $dX^i/dt$ and $X^i$ in these directions go to \textit{zero}, so the solutions in these cases are linear: $X^i \simeq at + b$. In particular, nearby geodesics will not re-intersect!

This leaves open the question of deviations about the \textit{blue} geodesic in Fig. \ref{fig:example-geodesics} -- the geodesic pointing towards the origin $z_2 = z_3 = 0$. In this case, we will have 
\begin{equation} z^i(t) \simeq (a, bt^{2/3}, ct^{2/3}), \q \frac{dz^i}{dt} \simeq \left(0, \frac{2}{3} bt^{-1/3}, \frac{2}{3} c t^{-1/3} \right)  \label{eq:bluegeosolution} \end{equation}
as $t \to 0$, and we find that
\begin{equation} \tensor{M}{^i_k}(t) = \frac{2}{3t} \mt{ 0  & & \\ & 1 & \\ & & 1 }, \q \tensor{N}{^i_l}(t) = - \frac{2}{9t^2} \mt{ 0  & & \\ & 1 & \\ & & 1 }.   \end{equation}
Now, there is a direction orthogonal to the trajectory in which the variation field $X(t)$ evolves nontrivially as follows: 
\begin{equation} \frac{d^2 X}{dt^2} + \frac{2}{3t} \frac{dX}{dt} - \frac{2}{9t^2} X = 0, \label{eq:deviationevolution} \end{equation}
where $\a = 2/3$. Substituting the new variable $\tau = \log t$, we find that
\begin{equation} \frac{d^2 X}{d\tau^2} - \frac{1}{3} \frac{dX}{d\tau} - \frac{2}{9} X  = 0.  \end{equation}
The discriminant of this linear second-order differential equation is given by
\begin{equation} \Delta =1 > 0.  \end{equation}
Thus, the solutions to this equation are exponential, not oscillatory. We conclude that the nearby geodesics must not re-intersect! 

It is in fact possible to exactly solve the geodesic deviation equation \eqref{eq:deviationevolution}:
\begin{equation} X(t) = C_1 t^{2/3} + C_2 t^{-1/3}. \end{equation}
In the case where $C_2 = 0$, the deviation is zero at the singularity $z_2 = z_3 = 0$. As $t$ increases, we see that the deviation $X(t)$ we see that $X(t)$ simply shifts the path $z(t)$ to approach the origin along a different direction at the same rate $t^{2/3}$. This makes good physical sense --- the coefficients $b, c$ in the expression for $z(t)$ in Eq. \eqref{eq:bluegeosolution} are arbitrary, so the geodesic deviation allows us to shift to different values of $b, c$. 

For $C_2 \neq 0$, we can engineer a solution $X(t)$ that satisfies $X(t_0) = 0$ for some $t_0 > 0$. These geodesics start at the same point at $t_0$, and $X(t) \to \infty$ as $t \to 0$. This tells us that nearby geodesics actually diverge from the original to the point that the assumption of infinitesimal deviation is no longer valid. Although this seems surprising, especially given the positive curvature, it actually represents the fact that an infinitesimal deviation from geodesic passing towards $(a, 0, 0)$ will instead hit the singular boundary at some $(a, \e, 0)$ or $(a, 0, \e)$ for $\e \neq 0$ before the deviation field can blow up.

We have explicitly demonstrated in this example that the geodesics behave perfectly regularly in the vicinity of the singular plane --- nearby geodesics do not re-intersect, and the geodesic deviation captures the different directions in which one may approach the singular locus. Although we have only illustrated the analysis for our chosen example, much more general and detailed arguments are given in Appendices \ref{sec:ex} and \ref{sec:dev}. 

From this view, the infinite distance limits with positive curvature become less mysterious: the curvature divergence at these singularities is related only to the geodesic incompleteness of the geometry, and the exponential map remains injective wherever it is defined. The attractive positive-curvature singularities at finite distance at the planes $y_2 = 0$ and $y_3 = 0$ correspond to cycles which reach zero size with finite action, an entirely regular process in the M-theory picture. 

\section{Unique Shortest Paths near QFT points in the Moduli Space Interior}
\label{sec:cft}

Thus far, we have studied geodesics near the asymptotic regions of marked moduli spaces, providing evidence for our Conjecture \ref{conj2'}. However, our stronger Conjecture \ref{conj2} pertains to geodesics \textit{anywhere} on marked moduli space, even in the interior. In this section, we will provide some evidence towards Conjecture \ref{conj2} from the moduli space interior. 

The connection between positive curvature in the interior of moduli spaces and points where a rigid QFT sector decouples from gravity has long been known. In the case of fully rigid supersymmetry, it was shown in \cite{caorsi_geometric_2018} that the Ricci curvature of the Coulomb branch geometry in the vicinity of the SCFT point is nonnegative. This behavior is observed to hold even in CY moduli spaces in the vicinity of decoupled QFT points, the prototypical examples of which are conifold transition points \cite{candelas_pair_1991}. 

In fact, this was the initial motivation for the \textit{asymptotic} negative curvature condition in \cite{ooguri_geometry_2007}, since it was already known from examples in \cite{candelas_pair_1991} that there exist isolated points at finite geodesic distance where the curvature diverges to $+\infty$. Our claim with Conjecture \ref{conj2} is that even near such decoupled QFT points, the unique shortest path property (a refinement of the unique geodesic property) will hold on the marked moduli space. In Section \ref{sec:conifold}, we will study this in the particular case of the conifold point for the mirror quintic CY3. In Section \ref{sec:othercft}, we will make some preliminary remarks about the behavior of geodesics in the vicinity of other decoupled QFT points.

It is also important to note that even if a counterexample were found to our conjectures in the vicinity of a finite-distance QFT point (though we do not know of any presently), we would still have our Conjecture \ref{conj2'}, valid in the \textit{asymptotic} regime. It is this statement about the asymptotic regime which still updates the asymptotic negative curvature conjecture of \cite{ooguri_geometry_2007}. Nevertheless, we still believe the stronger statement of Conjecture \ref{conj2} to be true, as we illustrate with the examples below. 

\subsection{The Conifold Point} \label{sec:conifold}

A prototypical example of divergent positive curvature is the case of a conifold point in the moduli space interior. This point corresponds to the well-studied conifold transition, where the volume of a three-cycle on a CY3 shrinks to zero. Deformations away from the conifold point correspond to giving finite volume to the three-cycle. 

Suppose that we have a single complex structure modulus $z$, parametrizing a moduli space with a conifold singularity. At the conifold point, the metric blows up as 
\cite{candelas_pair_1991}
\begin{equation} g_{z\bar{z}} \simeq -a^2 \log r, \end{equation}
where $r = \abs{z}$ is the distance from the conifold point and $a$ is a fixed positive constant. From this, it is seen that the scalar curvature diverges as 
\begin{equation} R \simeq \frac{A }{r^2 (-\log r)^3} . \end{equation}
for $A$ a positive constant. 

Given the divergent positive curvature at the conifold point, it seems \textit{a priori} possible that nearby geodesics could re-intersect near the conifold point in the marked moduli space. We wish to argue that this is not in fact the case. We will consider the case of geodesics pointing directly towards the conifold and study infinitesimal deviations from these geodesics. Using the geodesic deviation equation \cite{carroll_spacetime_2004}, we will see that nearby geodesics do not re-intersect before the conifold point is reached. 

Consider a geodesic pointing directly towards the conifold point $r = 0$, and suppose that our geodesic is parametrized by $t$ such that the point $r = 0$ is reached at $t = 0$. 
Now, we can always choose the parameter $t$ such that the geodesic moves at constant speed. The appropriate condition for such a constant-speed parameter $t$ is
\begin{equation}  -a^2 \log r \pfc{dr}{dt}^2 = \m{const}., \end{equation}
which implies that
\begin{equation} r \simeq \frac{t}{\sqrt{-\log t}}, \q t \to 0. \end{equation}
We now write down the geodesic deviation equation in two dimensions. In what follows, our discussion will be purely schematic; as we will see, we will not need to concern ourselves with overall factors. Furthermore, we concentrate solely on the coefficient of the deviation $X$, as this will be what controls whether the solution $X(t)$ is exponential or oscillatory. For our unit-speed geodesic, we find that a deviation $X$ obeys
\begin{equation} \frac{d^2 X}{dt^2} = - b R X \simeq - \frac{C}{ (- t \log t)^2} X \end{equation}
as $t \to 0$, where $b$ is another positive constant and $C = Ab$.

Now, this might initially seem problematic, since it appears \textit{a priori} that a divergent negative coefficient of $X$ could cause a nonzero geodesic deviation to collapse to zero before the point $t = 0$ is reached. However, let us study the following equation:
\begin{equation} \frac{d^2 X }{dt^2} = -\frac{\a^2}{4t^2} X . \end{equation}
In the variable $\tau = \log t$, we have
\begin{equation} \frac{d^2 X}{d\tau^2} - \frac{dX}{d\tau} + \frac{\a^2}{4} X = 0. \end{equation}
The discriminant of this linear second-order differential equation is 
\begin{equation} \Delta = 1 - \a^2. \end{equation}
Thus, for $\a^2 \leq 1$, the solutions are exponential and not oscillatory, and the geodesics will not re-intersect. Analogously, consider the equation
\begin{equation} \frac{d^2 X}{dt^2} = -\frac{f(t)^2}{4t^2} X. \end{equation}
For $f(t) \leq 1$, a solution for $X$ will not re-intersect $X = 0$ before $t = 0$. 

Going back to our example, the coefficient of $X$ diverges more slowly than $\a^2/(4t^2)$ for any $\a > 0$: For any $\a > 0$, and in particular for $\a = 1$, there will exist $\delta$ such that $\log^2 t > 4C/{\a^2}$ for $t < \delta$. Thus, by the argument in Section \ref{sec:unique}, we conclude that the deviation $X$ does not cross $X = 0$ arbitrarily close to the point $r = 0$. 

\begin{figure}
    \centering
\includegraphics[scale=0.6]{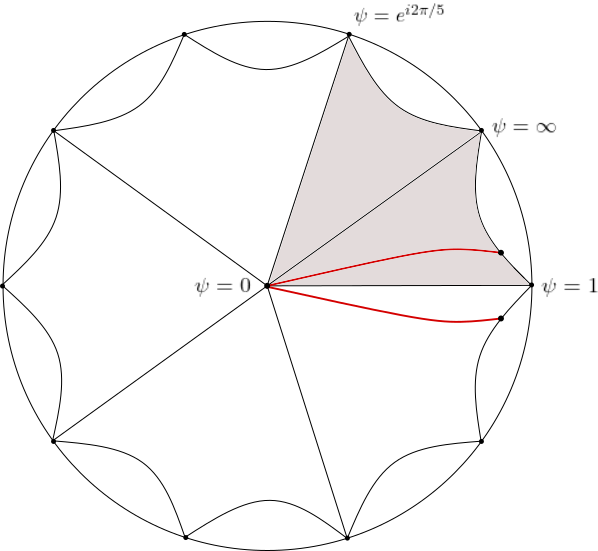}
    \caption{An illustration of the moduli space of the mirror quintic threefold as a quotient of the unit disk by the monodromy group. The marked moduli space $\hat{\C{M}}$ is the entire circle, while the unmarked moduli space $\C{M}$ is the fundamental region under the monodromy action, shaded in gray. Note that this picture does not represent distances --- the point $\psi = 1$ lies at finite distance in the moduli space, while $\psi = \infty$ lies at infinite distance. 
    Two geodesics emanating from $\psi = 0$ and passing close to the conifold $\psi = 1$ are indicated in red. Note that these geodesics re-intersect in the unmarked moduli space along the boundary from $\psi = 1$ to $\psi = \infty$, but they pass into different fundamental regions and do not re-intersect in the marked space.}
    \label{fig:conifold-fundamental}
\end{figure}

A concrete manifestation of the conifold point in a compact CY geometry arises in the case of the quintic threefold, first described in detail in \cite{candelas_pair_1991}. Here, the conifold point is a finite-distance curvature singularity in the complex-structure moduli space of the mirror quintic threefold. This example has a single complex-structure modulus $\psi$, and the metric $g_{\psi\bar\psi}$ has been computed exactly in terms of $\psi$. There are three fixed points of the monodromy group action. The Landau-Ginzburg point $\psi = 0$ is a fixed point of the $\B{Z}_5$ action $\psi \to e^{i2\pi/5} \psi$ has finite order; the large complex-structure point $\psi = \infty$ lies at infinite distance and has infinite order, while the conifold point $\psi = 1$ lies at \textit{finite} distance but also has infinite order. The global moduli space geometry is illustrated in Fig. \ref{fig:conifold-fundamental}.

In the quintic example, we will now argue more strongly that geodesics near the conifold do not re-intersect in the marked moduli space. Indeed, we understand the marked moduli space in the vicinity of the conifold point as follows. The conifold is a fixed point of a nontrivial $\B{Z}$ monodromy action. Now, specifying a basis of observables in the effective theory corresponds to specifying a basis for the periods of the CY3. But the monodromy group acts faithfully on the periods, so there is a period which transforms by $\B{Z}$ upon circling the conifold point. The marked moduli space in the vicinity of the conifold point is therefore a covering of $\B{C}$ whose fiber is $\B{Z}$: Any geodesic winding around the conifold in the unmarked space would move into the next Riemann sheet in the marked moduli space. In the unmarked space, we therefore see that we can effectively consider a branch cut extending from the conifold point $\psi = 1$ along the real axis $\m{Re}(\psi) = 0$ to infinity. 

Now, geodesics in the vicinity of the conifold point were studied numerically in \cite{blumenhagen_refined_2018}. The schematic behavior of these geodesics is depicted in Fig. \ref{fig:conifold-geodesics}. In particular, geodesics emanating from the Landau-Ginzburg point $\psi = 0$ pointing in a direction slightly away from the straight path to the conifold $\psi = 1$ were considered. These geodesics then ``curve'' around and re-intersect the axis $\m{Re}(\psi) = 0$. Thus, we see that geodesics do indeed re-intersect on the unmarked space --- one could draw two geodesics emanating from $\psi = 0$ at small initial angles $\pm \theta = \arg\psi(0)$.

\begin{figure}
    \centering
    \includegraphics[scale=0.5]{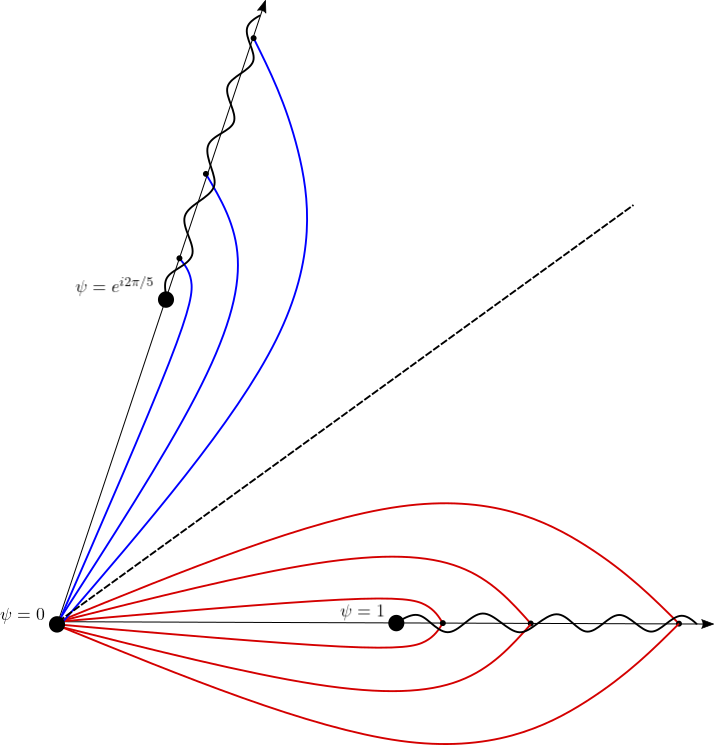}
    \caption{A schematic illustration of the geodesics calculated in \cite{blumenhagen_emergent_2023} (c.f. Fig. 5 in the original reference). The red geodesics emanate from $\psi = 0$ and re-intersect each other at the boundary $\m{Re}(\psi) = 0$ after the conifold point $\psi = 1$. The blue geodesics are simply the lower red geodesics, translated by $\B{Z}_5$ so that they lie in the fundamental region $0 < \arg(\psi) < 2\pi/5$. Note the branch cut extending from $\psi = 1$ off to infinity along the line $\m{Re}(\psi) = 0$.}
    \label{fig:conifold-geodesics}
\end{figure}

However, it was shown that for any geodesic ray emanating from the Landau-Ginzburg point $\psi = 0$, the axis $\m{Re}( \psi ) = 0$ is crossed only \textit{after} the conifold point $\psi = 1$ \cite{blumenhagen_refined_2018}. These geodesics only re-intersect after crossing the branch cut, so in the marked moduli space, the geodesics pass into different Riemann sheets. Moreover, since the vicinity of the Landau-Ginzburg point $\psi = 0$ is a completely regular in the WP geometry, it is in some sense completely generic to consider geodesics beginning at $\psi = 0$ and passing near the conifold. This provides strong evidence that geodesics emanating from any point and passing near the conifold point can only re-intersect with nontrivial monodromy about the conifold point. In the marked moduli space, the geodesics do not re-intersect. 

An alternative picture of this is shown in Fig. \ref{fig:conifold-fundamental}. As shown in \cite{candelas_pair_1991}, the unmarked moduli space is a fundamental region of the monodromy group action on the unit disk. Nearby geodesics that re-intersect in the unmarked space on the line $\m{Re}(\psi) = 0$ after the conifold point $\psi = 1$ pass into entirely different fundamental regions and do not intersect in the marked moduli space. 

We have here argued for the uniqueness of geodesics, if they exist, between any two points in the marked moduli space of the mirror quintic. However, we note that this does \textit{not} imply that there exists a geodesic between any pair of points. Indeed, the failure of the marked moduli space $\hat{\C{M}}$ to be complete, as in the conifold example, is related to the observation\footnote{We are  grateful to Sergio Cecotti for his insights and detailed analysis in the case of the conifold that brought this point to our attention, leading to a proper restatement of our conjecture in cases with finite curvature singularities.} that there does not necessarily exist a geodesic between any pair of points in $\hat{\C{M}}$ with respect to the physical metric. 

To see what is going on, we first note that the asymptotic geometry in the vicinity of the conifold point is specified in polar coordinates by the following metric, up to an overall scaling: 
\begin{equation} ds^2 = -\log r(\d r^2 + r^2 \d \theta^2). \end{equation}
In these coordinates, the conifold is located at $r = 0$. From this, we find the following geodesic equations: 
\begin{align}
    \frac{d^2 r}{dt^2} + \frac{1}{2r \log r} \pfc{dr}{dt}^2 - r\left(1 + \frac{1}{2\log r} \right) \pfc{d\theta}{dt}^2 &= 0 \\
    \frac{d^2 \theta}{dt^2} + \frac{2}{r} \left(1 + \frac{1}{2\log r} \right)  \frac{dr}{dt} \frac{d\theta}{dt} &= 0. 
\end{align}
In the limit as $r$ approaches $0$ (such that $-\log r \gg 1$), we observe that the geodesic equations reduce (after a redefinition of the parameter $t$) to the geodesic equations for flat space in polar coordinates. Thus, closer to the conifold, the geodesics look more like straight lines. In particular, starting at a point $(r_0, 0)$, there is a finite angle $\hat{\theta} (r_0)$ such that there is a geodesic connecting $(r_0, 0)$ and $(r_0, \theta)$ iff $\abs{\theta} < \hat{\theta} $. In practice, $\hat{\theta} \simeq \pi +\delta(r_0)$ for $-\log r_0 \gg 1$ where $\delta(r_0)\rightarrow 0$ as $r_0\rightarrow 0$. In particular, geodesics cannot wind around the conifold point arbitrarily many times, as would be required to reach points in Riemann sheets arbitrarily far away in the marked moduli space. Thus, we do not expect there to be a geodesic connecting an arbitrary pair of points in the marked moduli space. 

However, Conjecture \ref{conj2} posits the \textit{unique shortest path property}, a technical refinement to the unique geodesic property, and this property still appears to hold in the mirror quintic moduli space. In particular, we have already provided evidence for the uniqueness of geodesics, if they exist, between any two points. Although a geodesic does not necessarily exist between any two points $p, q$ in $\hat{\C{M}}$, it is still possible to construct a ``canonical'' shortest path between $p, q$ as follows. Suppose $p, q$ are not connected by a geodesic. We note that the conifold point is reachable by both $p, q$ via a unique geodesics $\g_p, \g_q$. Then the canonical shortest path between $p, q$ is the two-segment path that passes through the conifold by concatenating $\g_p$ and $\g_q$. This setup is illustrated in Fig. \ref{fig:conifold-geodesics}. We see that this mechanism works since it is possible to traverse an arbitrary angle in $\hat{\C{M}}$ in zero distance by reaching the conifold point. By the same mechanism, we therefore conjecture that the unique shortest path property in fact holds for all marked moduli spaces with finite-distance curvature singularities. 

\begin{figure}
\centering
\begin{subfigure}{\textwidth}
  \centering
  \includegraphics[width=0.8\linewidth]{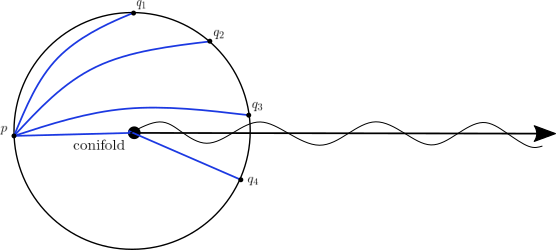}
  \label{fig:conifold-geodesics-unmarked}
\end{subfigure}
\begin{subfigure}{\textwidth}
  \centering
  \includegraphics[width=0.8\linewidth]{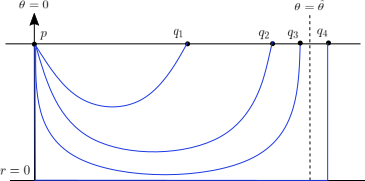}
  \label{fig:conifold-geodesics-marked}
\end{subfigure}
\caption{An illustration of various geodesics and non-geodesic paths between pairs of points in the vicinity of the conifold. Starting from point $p$, the shortest paths to $q_1, q_2, q_3$ are the unique blue geodesics shown. However, there is no geodesic between $p$ and $q_4$, since the angular excursion from $p$ is greater than the finite upper bound $\hat{\theta}$ that geodesics from $p$ may traverse. Thus, the unique shortest path is instead a piecewise-geodesic path that passes directly into the conifold and ``pops out'' in a different direction. The top figure shows this setup in the unmarked moduli space, while the bottom figure shows this setup in the marked moduli space. Note that the locus $r = 0$ in the marked moduli space should be identified as a single point, so that it takes zero distance to jump between the two segments of the path from $p$ to $q_4$. As $\theta \to \hat{\theta}$ from below, the geodesic paths approach the singular piecewise path.}
\label{fig:conifold-geodesics}
\end{figure}

\subsection{Argyres-Douglas Points} \label{sec:othercft}

Motivated by the conifold example, we briefly consider geodesics in the vicinity of other QFT points in CY3 moduli spaces. We will focus on the case of maximal Argyres-Douglas (AD) points with gauge group $\SU(3)$ which arise from the compactification of M-theory on a CY3 geometry with an A-type singularity (as per the classification in  \cite{eguchi_n2_1996}). The 4d $\C{N}=2$ Argyres-Douglas SCFT lives at the AD point, and small perturbations away from this point in the CY3 moduli space correspond to moving along the Coulomb branch of the SCFT.

We will briefly review some basic facts about the Coulomb branch geometry of 4d $\C{N}=2$ SCFTs below. Let us consider CY3 singularities of Type A, given by 
\begin{equation} y^2 = x^{3} + uv . \end{equation}
A deformation about this singularity involves sending 
\begin{equation} x^3 \to P(x) = x^3 - g_0 .\end{equation}
On the other hand, the vicinity of maximal AD points of pure gauge $\C{N}=2$ super Yang-Mills theory with gauge group $\SU(N)$ is described by the Seiberg-Witten (SW) curve $y^2 = P(x)$, and the corresponding SW differential is
\begin{equation} \l = i y \d x. \end{equation}
The $g_i$ are then interpreted as the vevs of a basis of Coulomb branch operators. 

The Riemann surface $\Sigma$ defined by $y^2 = P(x)$ has genus one and admits a basis of cycles $A, B$ for its first homology. The periods of $\l$ around $A, B$ are given by 
\begin{equation} a = \frac{1}{2\pi i} \oint_{A} \l , \q a_{D} = \frac{1}{2\pi i} \oint_{B} \l . \end{equation}
The Coulomb branch geometry in the vicinity of these AD SCFT points corresponds is specified by a Kähler metric whose Kähler form is. 
\begin{equation}  \w = -\frac{i}{4\pi} \left(\d a \wedge \d \bar{a}_{D} - \m{c.c.} \right)   \end{equation}
Since the SW periods $a$ and $a_{D}$ have the interpretation of BPS masses in the theory, the SW differential has scaling dimension one. From this, we conclude that the Coulomb branch operator dimension $[g_0]$ is given by
\begin{equation} \Delta := [g_0] = \frac{6}{5} . \end{equation}

We now see that the metric is flat away from the AD point, where there is a conical singularity. In any case, nearby geodesics approaching $g_0 = \bar{g}_0 = 0$ will not re-intersect. A simple dimensional argument tells us that the Kähler form is given by 
\begin{equation} \omega \simeq \abs{g_0}^{-1/3} \d g_0 \wedge \d\bar g_0. \end{equation}
The curvature is then
\begin{equation} R_{i\bari} = -\pd{}{g_0} \pd{}{\bar{g}_0} \log \abs{g_0}^{-1/3} = \frac{2\pi}{6} \delta^2(g_0).
\end{equation}
The metric is flat everywhere except at the cone point $g_0 = 0$. In effect, the geometry is what is obtained by ``folding'' a wedge of the complex plane with $0 \leq \arg(z) \leq \frac{5}{6}(2\pi)$. 

It appears at first that one might be able to consider a nontrivial geodesic that winds around the conical singularity and re-intersects itself in the unmarked moduli space. Here, we argue that by uplifting to the marked moduli space, such geodesics no longer self-intersect.

To begin, we first must understand what a notion of marking corresponds to on this moduli space. A basis of observables is specified by the BPS lattice of central charges, which is generated by the periods $a, a_D$. Indeed, we can compute $a, a_D$ in terms of $g_0$: 
\begin{equation} a \simeq  g_0^{5/6}, \q a_D \simeq \bar{g}_0^{5/6}. \end{equation}
But we now notice that $a, a_D$ transform by a monodromy as we wind around the conical singularity at $g_0 = 0$. Thus, in the marked moduli space, a geodesic path winding around the singularity actually moves into a different Riemann sheet. The map from the marked moduli space to the unmarked moduli space therefore apparently corresponds to the sixfold covering map $p : z \mapsto z^6$. 

However, this sixfold covering is not exactly the marked moduli space. Indeed, letting $f_0 = g_0^{1/5}$, note that the observables $a, a_D$ satisfy $a \simeq f_0^5$ and $a_D \simeq \bar{f}_0^5$. Thus, the $a, a_D$ actually exhibit a fivefold redundancy relative to the $f_0, \bar{f}_0$. As such, the true marked moduli space is a fundamental region of this $\B{Z}_5$ action on the full sixfold covering space parametrized by the $f_0, \bar{f}_0$ variables. We therefore see that the marked moduli space that the physical observables $a, a_D$ can distinguish is in fact just the flat space $\B{C}$, including the point at the origin (which no longer has a conical singularity). Thus, one can see in fact that the unique geodesic property holds, which certainly implies Conjecture \ref{conj2}. In this way, we see that the \textit{marking} on moduli space is what saves us --- the labeling of observables in the theory is crucial for the Conjecture \ref{conj2} to hold. Indeed, this procedure works in a very nontrivial way for this AD example, lifting the conical singularity with deficit angle $2\pi/6$ to the flat space $\B{C}$ with no conical singularity.

It would be interesting to study the metric and geodesics in the vicinity of more complicated AD points, which are known to have positive Ricci curvature \cite{caorsi_geometric_2018}. It was argued in \cite{bharadwaj_approaching_2023} that the Kähler potential for the Coulomb branch moduli space is convex in a small neighborhood of AD points, and explicit expressions for the Kähler potential were given for several examples. Although these formulas are typically complicated, and studying geodesics in the entire Coulomb branch geometry might prove difficult, it may be possible to do a local analysis of neighboring geodesics in the vicinity of the QFT point, as we have done above for our simple examples.

\section{Consequences of the Conjectures} \label{sec:concl}

In this work, we have studied the geometry and topology of marked moduli spaces in quantum gravity. Motivated by a diverse array of supersymmetric examples, we have conjectured that marked moduli spaces are contractible and moreover that they satisfy the unique geodesic property with respect to the physical metric (arising from the EFT description). In this section, we will discuss some physical motivations for and implications of our conjectures. 

Let us first examine our topological Conjecture \ref{conj1} -- the contractibility of marked moduli spaces. The contractibility of configuration spaces associated to quantum gravity is indeed a prediction of the Cobordism Conjecture (CC).  However, this occasionally requires us to go over potential barriers.  Here we have focused on the vacua with minimum potential, so our conjecture is only morally related to the CC. One might ask why we are forced to consider the \textit{marked} moduli space instead of the moduli space itself to get the desired properties. However, the spectrum of the theory will generically be acted on by the duality group. By choosing a basis for a set of observables transforming in a nontrivial representation of the duality group, we therefore pass to a covering space of the moduli space $\C{M}$. The statement of completeness of the quantum gravity spectrum tells us that there should always exist objects transforming in a faithful representation of the duality group, in which case a choice of basis forces us to pass to the \textit{universal} cover of the moduli space $\C{M}$.  Thus, precisely due to the completeness hypothesis, the marked moduli space $\hat{\C{M}}$ is a well-defined and natural space with which to parametrize the vacua of the theory. 

Let us now consider our \textit{geometric} Conjecture \ref{conj2} -- the unique shortest path property of marked moduli spaces. Firstly, and perhaps most straightforwardly, the unique shortest path property allows us to unambiguously define a distance function on a marked moduli space $\hat{\C{M}}$. In particular, the distance between $p, q \in \hat{\C{M}}$ is defined to be the length of the \textit{unique} shortest path between $p$ and $q$. If the unique shortest path condition did not hold, one could still define a notion of distance on $\hat{\C{M}}$ by choosing the distance between points to be the length of one of the shortest paths between $p$ and $q$; however, this is less natural than if there were no ambiguity in the choice of path (as we are proposing).

Furthermore, recall that the CC tells us of the existence of a defect interfacing between any two consistent vacua of quantum gravity.   Our conjectures would imply that there is a unique action-minimizing configuration, \textit{up to duality}, between any two vacua $p, q \in \hat{\C{M}}$ corresponding to the shortest path between $p, q$.  This makes this notion of a canonical domain wall between vacua more elegant.

Another consequence of our conjectures has to do with \textit{closed geodesics} -- geodesics that start and return to the same point with the same tangent vector. The unique shortest path property posits the uniqueness of geodesics, if they exist, between any two points, and this therefore forbids the existence of closed geodesics on marked moduli space. Thus, there are no closed geodesics on the \textit{unmarked} moduli space $\C{M}$ without a nontrivial action by the duality group. In fact, one can say more: On $\C{M} = \Gamma\backslash\hat{\C{M}}$, every closed geodesic lies in one-to-one correspondence with a conjugacy class of the duality group $\Gamma$. Looking at this from the perspective of time evolution, our conjectures therefore seem to imply that, unlike the ``bouncing cosmology'' scenarios, the only bouncing solutions that are allowed are ``duality bounces'' which undergo some duality transformation from one cycle to the next. This is indeed the case in the string gas cosmology of \cite{brandenberger_superstrings_1989}, where a T-duality transformation relates one cycle to the next. 

A further interpretation for the absence of closed geodesics on marked moduli space is as follows. Suppose we compactify a theory with moduli space $\C{M} = \Gamma\backslash\hat{\C{M}}$ on a circle $S^1$. Then the nontrivial cobordism classes in dimension 1 are generated by circles $S^1$ equipped with a nontrivial bundle for the duality group $\Gamma$. Thus, there is a nontrivial defect of codimension two associated to a conjugacy class of $\Gamma$. In the context of our conjectures, we have a much stronger statement: There is a unique action-minimizing configuration for each cobordism defect trivializing $S^1$ with a nontrivial duality bundle. It is in this sense that our conjectures represent a ``geometrization'' of the topological CC in the vacuum sector. More generally, our conjectures speak to a larger pattern of ``topology to geometry'' features of quantum gravity\footnote{In a sense, this is similar to the relation between the no-global-symmetries conjecture (the analog of topology) and the WGC (the analog of geometry).}.
The defects predicted by the vacuum version of CC on topological grounds must have finite tension, which should in principle be computable and related to the topological charge of the defect. For example, it would be interesting to compute the lengths of the closed geodesics on unmarked moduli spaces and relate them to their associated duality charges. More broadly, we believe that it will be important to shed further light on the deep relationship between topological and geometric properties of quantum gravity. 
Finally, in this paper, we have focused on the vacuum sector of the theory. It would be important to try to generalize our conjectures to the case of scalar fields with potentials. This will bring our conjectures in more contact with the setup relevant for the Cobordism Conjecture.

\subsubsection*{Acknowledgments} 
We would like to thank Sergio Cecotti for sharing his insights on geometry of the ${\mathcal N}=2$ moduli space.  We have also benefited from discussions with Jake McNamara, Miguel Montero and Hirosi Ooguri.

This work is supported in part by a grant from the Simons Foundation (602883,CV), the DellaPietra Foundation, and by the NSF grant PHY-2013858.

\appendix

\section{General Consideration of Geodesics in the Asymptotic Regime} \label{sec:prel}

In this section, we will write down the geodesic equation on the Teichmüller space $\hat{\C{M}}$ in the asymptotic large-volume regime. Recall that $\hat{\C{M}}$ is precisely the Kähler cone whose Kähler metric is specified by the prepotential given in Eq. \eqref{eq:lvpre}: 
\begin{equation} C = C_{ijk} y^i y^j y^k. \end{equation}
We define as usual
\begin{equation} C_{i} = C_{ijk} y^j y^k, \q C_{ij} = C_{ijk} y^k. \end{equation}
From this, we derive the metric: 
\begin{equation} g_{ij} = \frac{9}{4} \frac{C_i C_j}{C^2} - \frac{3}{2} \frac{C_{ij}}{C}. \end{equation}
At this point, since we are working only with the real variables $y^i$, it becomes convenient to work only in terms of these coordinates. We therefore drop the bars on indices henceforth. For reference, we also quote the inverse metric: 
\begin{equation} g^{ij} = 2z^i z^j - \frac{2}{3} C C^{ij}, \end{equation}
where $C^{ij}$ is the inverse to $C_{ij}$. 

It is shown in \cite{viaclovsky_kahler_2018} that the only nonzero Christoffel symbols are $\Gamma^i_{jk}$ and $\Gamma^{\bari}_{\barj \bar k}$. The Christoffel symbols are moreover given by
\begin{equation} \Gamma^i_{jk} = \frac{1}{2} g^{il} \p_j g_{k l}, \end{equation}
where the derivative is with respect to the real $y$-variable. Note that this formula
contains a factor of $\frac{1}{2}$ relative to what is sometimes seen in the literature; this is due to the fact that the derivative with respect to the complex variable $x^i + iy^i$ contains a factor of $\frac{1}{2}$ relative to the derivative with respect to the real variable $y^i$. The Christoffel symbols are given by the following expression: 
\begin{equation} \Gamma^i_{jk} = \frac{1}{2} \left[ \frac{3 z^i C_{jk} - 3 \delta^i_j C_k - 3 \delta^i_k C_j }{C} + C^{il} C_{jkl} \right] . \end{equation}

The geodesic equation is
\begin{equation} \frac{d^2 y^i}{dt^2} + \Gamma^i_{jk} \frac{dy^j}{dt} \frac{dy^k}{dt} = 0. \end{equation}
Substituting in our expression for $g_{ij}$, and clearing some overall factors from all of the terms, we get
\begin{equation}
    \begin{gathered}
        \left[ 3 CC_i C_l - 2 C^2 C_{il} \right] \frac{d^2 y^i}{dt^2} + \\ \left[ 3C (C_j C_{kl} + C_k C_{db} + C_l C_{jk}) - 9 C_j C_k C_l - C^2 C_{jkl}  \right] \frac{dy^j}{dt} \frac{dy^k}{dt} = 0 . 
    \end{gathered}
\end{equation}
We first contract the geodesic equation against $y^i$: 
\begin{equation} C^2 C_i \frac{d^2 y_i}{dt^2} + (2C^2 C_{jk} - 3CC_j C_k) \frac{dy^j}{dt} \frac{dy^k}{dt} = 0.  \end{equation}
Notice now that the geodesic equation reads 
\begin{equation} \frac{1}{3} C^2 \frac{d^2 C}{dt^2} - \frac{1}{3} C \pfc{dC}{dt}^2 = 0. \end{equation}
which is solved by 
\begin{equation} C(t) = C_0 e^{\a t}. \end{equation}
Inserting this solution back in, we observe that
\begin{equation}  2 C^2 C_{il} \frac{d^2 y^i}{dt^2} + 3 C C_l C_{jk} \frac{dy^j}{dt} \frac{dy^k}{dt} + C^2 C_{jkl} \frac{dy^j}{dt} \frac{dy^k}{dt} = 2\a C^2 C_{il} \frac{dy^i}{dt} .\end{equation}
Contract this against $dy^d/dt$. Collecting terms and simplifying, we find that
\begin{equation} C \frac{d}{dt} \left[ C C_{il} \frac{dy^i}{dt} \frac{dy^d}{dt}  \right] = 2\a C \left[ C C_{il} \frac{dy^i}{dt} \frac{dy^d}{dt}  \right]. \end{equation}
We conclude that
\begin{equation}  C_{il} \frac{dy^i}{dt} \frac{dy^d}{dt} = g_0 C_0 e^{\a t}.  \end{equation}
for $g_0$ a constant. Note that
\begin{equation} g_{ij} \frac{dy^i}{dt} \frac{dy^j}{dt} = \frac{1}{4} \a^2 - \frac{3}{2} g_0 ,\end{equation}
so the geodesic moves with constant speed.

Substituting this condition in, we obtain the following equation: 
\begin{equation}  2  C_{il} \frac{d^2 y^i}{dt^2} - 2\a C_{il} \frac{dy^i}{dt}  + C_{jkl} \frac{dy^j}{dt} \frac{dy^k}{dt} = - 3 g_0 C_l .\end{equation}
Now, set
\begin{equation} y^i = w^i e^{\a t/3}. \end{equation}
Substituting in the new variable $w^i$, we conclude that
\begin{equation} C_{ijk} w^i w^j \frac{dw^k}{dt} = 0, \q C_{ijk} w^i \frac{dw^j}{dt} \frac{dw^k}{dt} = g_0 + \frac{\a^2}{9} := h_0 . \end{equation}
In particular, we observe that
\begin{equation} g_{ij} \frac{dw^i}{dt} \frac{dw^j}{dt} = -\frac{3}{2C} h_0 > 0,\end{equation}
implying that $h_0 < 0$. The final simplified form of the geodesic equation is 
\begin{equation} 2  C_{il} \frac{d^2 w^i}{dt^2}   + C_{jkl} \frac{dw^j}{dt} \frac{dw^k}{dt} = \l C_l, \label{eq:geofinal} \end{equation}
for $\l$ a positive constant, along with the constraint equation
\begin{equation} C_{ijk} w^i w^j w^k = C_0. \end{equation}
The geodesics are of the form $y^i = w^i e^{\a t/3}$, where the $w^i$ solve the above equations and $\a$ is an arbitrary real constant. Since all geodesic solutions $w^i$ differ from the solutions $z^i$ by overall scale factors $e^{\a t/3}$, it therefore suffices to study only the geodesics $w^i$ for which $\a = 0$. 

\section{Details of Example Asymptotic Geodesic Calculation} \label{sec:ex}

In this appendix, we will prove that the unique geodesic property (and therefore Conjecture \ref{conj2'}) holds \textit{everywhere} in the asymptotic Kähler moduli space for the example discussed in Section \ref{sec:geo} with the following prepotential: 
\begin{equation} C = z_1^3 - z_2^3 - z_3^3. \end{equation}
To begin, recall the general form of the geodesic solutions described in Section \ref{sec:geo}:
\begin{equation} z_i(t) = e^{\a t/3}(A_i e^{\b t} + B_i e^{-\b t})^{2/3}. \end{equation}
where $\a, \b$ are arbitrary real numbers and the $A_i, B_i$ are real numbers satisfying the constraints
\begin{equation} A_1^2 = A_2^2 + A_3^3, \q B_1^2 = B_2^2 + B_3^2 . \end{equation}

We wish to show that the unique geodesic condition holds over all such geodesics. To do this, we will pick an arbitrary starting point $p_i^{2/3} = z_i(0)$ and an arbitrary other point with coordinates $q_i^{2/3} = z_i(1)$. We will show that there is a unique geodesic along which one must travel unit time to reach an arbitrary point $y_i(1)$ from $p$. Note that
\begin{equation} z_i(1) =  e^{\a /3}(A_i e^{\b } + B_i e^{-\b })^{2/3} . \end{equation}
Write 
\begin{equation} p_i = A_i + B_i, \q q_i = e^{\a/2}(A_i e^{\b } + B_i e^{-\b }). \end{equation}
We will choose the starting $A_i, B_i$ such that the $p_i > 0$. Moreover, we will also insist that the $q_i > 0$. This ensures that we will not go through the singularity $z_i = 0$ as we pass from $z_i(0)$ to $z_i(1)$. 

Finding the appropriate $A_i, B_j, \a, \b$ amounts to solving the following eight equations: 
\begin{equation} p_i = A_i + B_i, \q q_i = a A_i + b B_i, \q A_1^2 = A_2^2 + A_3^3, \q B_1^2 = B_2^2 + B_3^2. \end{equation}
Here, we have defined $a = e^{\a/2+\b}, b = e^{\a/2 - \b}$; these are arbitrary positive real constants. We aim to show that the above set of equations yields a single geodesic. To show this, we use a few geometric tricks. Note that the $A_i$ lie at the intersection of two cones $N$ and $P$: 
\begin{equation} N : A_1^2 = A_2^2 + A_3^2, \q  P : (p_1 - A_1)^2 = (p_2 - A_2)^2 + (p_3 - A_3)^2. \end{equation}
Since $p_i$ lies in the interior of the cone in the first quadrant, the intersection of these two cones will be an ellipse; this ellipse defines the allowed values for the vector $A_i$. 

The $A_i$ must now also be made to lie on the cone $Q$ defined by 
\begin{equation} Q: (q_2 - aA_1)^2 = (q_2 - aA_2)^2 + (q_2 - a A_3)^2, \end{equation}
so the allowed $A_i$ lie at the intersection of the ellipses defined by the intersection of the intersection of these cones $P, Q$ with $N$. Moreover, the vector difference between $aA_i$ and $q_i$ must be proportional to $B_i$, which is equal to $p_i - A_i$. This means that by going along the line from $A_i$ to $p_i$, we must arrive at $q_i$. In particular, the vertex of $Q$ must lie on $P$, so the cones must be \textit{tangent}. This yields the following quadratic equation in $a$:
\begin{equation} (a p_1 - q_1)^2 = (ap_2 - q_2)^2 + (ap_3 - q_3)^2 . \end{equation}
The discriminant of this equation is 
\begin{equation} 4 \left[ (p_1 q_1 - p_2 q_2 - p_3 q_3)^2 - (p_1^2 - p_2^2 - p_3^2)(q_1^2 - q_2^2 - q_3^2) \right], \end{equation}
which is positive by the ``Lorentzian'' Cauchy-Schwarz inequality. Moreover, the roots of the quadratic equation are also positive, since $p_1^2 - p_2^2 - p_3^2 > 0$ and $q_1^2 - q_2^2 - q_3^2 > 0$. We therefore have two possible values for $a$. These fix $A_i$ as the tangency point of the two ellipses formed by the intersection of the tangent cones $P, Q$ with $N$. Finally, $b$ is fixed as the ratio of the distance from $A_i$ to $q_i$ vis-à-vis the distance from $A_i$ to $p_i$. 

It therefore seems like there are \textit{two} geodesics. But this is illusory: the two solutions are actually equivalent after the interchange $A_i \leftrightarrow B_i$ and $a \leftrightarrow b$. We therefore conclude that there is a unique geodesic along which one must travel for a unit time (without encountering a singularity) to go from $z_i(0)$ to $z_i(1)$. This shows that the exponential map is injective, exactly as required. Note that the Kähler cone is not geodesically complete, of course --- there are geodesics which reach the singularity at finite time.

\section{Geodesic Deviation near Arbitrary Asymptotic Positive-Curvature Limits} \label{sec:dev}

In this appendix, we will provide a general argument that the unique geodesic property holds in the vicinity of a region with divergent positive curvature in \textit{any} Kähler cone geometry with a cubic prepotential $C = C_{ijk} y^i y^j y^k$. To be precise, we will show that nearby geodesics do not re-intersect arbitrarily close to a locus with divergent positive curvature. This provides strong evidence that the unique shortest path property holds everywhere --- although we do not exclude the possibility that geodesics could re-intersect somewhere at finite distance from the curvature singularity, this seems much more unlikely. We will focus on geodesics in the $w^i$ variables at which the volume $C$ is held fixed, since all other geodesics are related by a simple scaling by $e^{\a t/3}$ (as argued in Appendix \ref{sec:prel}). 

We wish to show that the $w^i$ obey the unique geodesic property despite the possibility of a finite-distance positive curvature singularity locus $\C{R}$ at which $h = \det C_{ij} = 0$. In the M-theory picture, such geodesics correspond to paths along which the volume of a chosen divisor of the Calabi-Yau shrinks to zero.
\begin{equation} v^i C_i = v^i C_{ijk} w^j w^k  \to 0 . \end{equation}
This corresponds to approaching a decoupled QFT point in the physical theory, as explained in \cite{marchesano_moduli_2023}. At the singular point $w_0 \in \B{R}$, the matrix $C^0_{ij} = C_{ijk} w_0^k$ has reduced rank, and $v^i \in \ker C^0_{ij}$. 

We will consider an arbitrary geodesic $\g$ emanating from a point $p$ sufficiently close to $\B{R}$ that pass towards $\C{R}$ (in the direction of diverging scalar curvature) along precisely a path as described above. We will study the geodesic deviation equation about this geodesic and show that $p$ has no conjugate points where nearby geodesics re-intersect along $\g$. 

First, recall the geodesic deviation equation \eqref{eq:geodev2}: 
\begin{equation} \frac{d^2 X^i}{dt^2} + 2\Gamma^i_{jk} \frac{dw^j}{dt} \frac{dX^k}{dt} + \p_{l} \Gamma^i_{jk} \frac{dw^j}{dt} \frac{dw^k}{dt} X^{l} = 0 . \label{eq:geodevapp} \end{equation} 
The goal is to study the singular pieces of the coefficients in the geodesic deviation equation as the singularity is approached. Let us now study geodesics in the vicinity of the positive curvature divergence. That is, we will look at geodesic paths which send the volume of a 2-cycle in the CY geometry to zero. Since the singularity is approached in finite geodesic distance, we will take the time at which the singularity is reached to be $t = 0$. Generically, we will write the expression for these geodesics as
\begin{equation} w^i(t) = w^i_0 + e_1^i t^{\a_1} + e_2^i t^{\a_2} + \cdots, \label{eq:expansion} \end{equation}
where $\a_1 <  \a_2 < \cdots$ and where $z^i_0$ is a point at which $C_{ij}$ has a zero eigenvalue (equivalently, where $h = \det C_{ij} = 0$). We then have
\begin{equation} \frac{dw^i}{dt} = \a_1 e_1^i t^{\a_1 - 1} +  \a_2 e_2^i t^{\a_2 - 1} + \cdots \end{equation}

Let us now determine the possible values of $\a_i$.  We begin with the final version Eq. \eqref{eq:geofinal} of the geodesic equation derived in Appendix \ref{sec:prel}:
\begin{equation} 2  C_{il} \frac{d^2 w^i}{dt^2}   + C_{jkl} \frac{dw^j}{dt} \frac{dw^k}{dt} = \l C_l,  \end{equation}
where $\l$ is a positive constant. We also have the constraint
\begin{equation} C_{ijk} w^i w^j \frac{dw^k}{dt} = 0. \end{equation}
Now, substitute Eq. \eqref{eq:expansion} into the geodesic equation. We make the \textit{ansatz} that $\a_1 \neq 1$. We collect terms proportional to each power of $\a_i$. The leading term arises from the $d^2 w^i/dt^2$ piece alone and goes as $t^{\a_1 - 2}$: 
\begin{equation} 2\a_1 (\a_1 - 1) t^{\a_1 - 2} C_{ijl} w_0^j e_1^l  = 0.  \end{equation}
We therefore demand that $C_{ijl} w_0^j e_1^l = 0$. In particular, $e_1^l \in \ker g_{ij}$. But note that the singular locus is \textit{defined} by the fact that $\ker g_{ij} \neq 0$. Moreover, the vector $e_1^l$ is not parallel to $w_0^l$ since we are at finite volume (and therefore $C_{ijk} w_0^i w_0^j \neq 0$). Thus, a generic geodesic path approaching the singularity will \textit{always} feature a component along $e_1^i$ for which $C_{ijl} w_0^j e_1^l = 0$. We can determine the corresponding exponent $\a_1$ by checking the coefficient of $t^{2\a_1 - 2}$. Note that the right-hand side is forced to be zero at this order in $t$ by the condition $C_{ijl} w_0^j e_1^l = 0$. We therefore find that
\begin{equation} \left[ 2\a_1(\a_1 - 1) + \a_1^2 \right] (C_{ijk} e_1^j e_1^k) = 0,  \end{equation}
implying that 
\begin{equation} \a_1 = \frac{2}{3}. \end{equation}
We have found that the generic leading behavior of a path approaching the curvature singularity in \textit{any} special Kähler geometry is given by 
\begin{equation} w^i(t) = w_0^i + e^i t^{2/3} + \cdots, \end{equation}
where $C_{ijk} w_0^i e^j = 0$. This precisely aligns with the example calculated in Section \ref{sec:geo}. In fact, it is this key feature which allows us to show that the geodesics in the vicinity of the singularity do not re-intersect.

We now return to the geodesic deviation equation. Recall that we computed the leading divergent contribution to the Christoffel symbols in Section \ref{sec:geo}:
\begin{equation} \Gamma^i_{jk} \simeq \frac{1}{2} C^{il} C_{jkl} + \cdots  \end{equation}
and
\begin{equation} \p_l \Gamma^i_{jk} \simeq -\frac{1}{2} C^{im} C_{mnl} C^{np} C_{jkp} + \cdots. \end{equation}
Let us now consider the following matrix: 
\begin{equation} \tensor{M}{^i_j} := C^{il} C_{jkl} \frac{dw^k}{dt} = \frac{2}{3} t^{-1/3} C^{il} C_{jkl} e^k + \cdots   \end{equation}
We wish to determine the most singular eigenvalues of this matrix. We first note that the most singular eigenvalues of $C^{ij}$ lie precisely along the directions which are sent to zero by $C_{ij}$. Thus, eigenvectors which are not in the kernel of $C_{ijk}z_0^k$ are acted on perfectly regularly by $C^{ij}$. We see therefore that the corresponding eigenvalues diverge at most as badly as $t^{-1/3}$. As we will argue later, such a divergence is not bad enough to allow the re-intersection of geodesics arbitrarily close to the singularity.

For now, let us restrict our attention to vectors annihilated by $C_{ijk} w_0^k$. Suppose that we are given such a vector $v^j$. Then we have
\begin{equation} \tensor{M}{^i_j} v^j = \frac{2}{3} t^{-1/3} C^{il} C_{jkl} e^k v^j  = \frac{2}{3t} C^{il} C_{jkl}(w_0^k + e^k t^{2/3} )v^j \simeq \frac{2}{3t} C^{il} C_{jl} v^j = \frac{2}{3t} v^j. \end{equation}
We therefore see that every vector $v^k \in \ker C_{ij}$ is an eigenvector of $\tensor{M}{^i_j}$ with eigenvalue $2/(3t)$. 

Using this, we may simplify the expression for $\p_l \Gamma^i_{jk}$ as follows: 
\begin{equation} \p_l \Gamma^i_{jk} \frac{dw^j}{dt} \frac{dw^k}{dt}  \simeq -\frac{2}{9} t^{-2/3} C^{im} C_{mnl} C^{np} C_{jkp} e^j e^k . \end{equation}
Noting that $e^k \in \ker C_{kp}$, we find that
\begin{equation}- \frac{2}{9} t^{-2/3} C^{np} C_{jkp} e^j e^k -\simeq -\frac{2}{9} t^{-4/3} e^n,  \end{equation}
so that
\begin{equation} \p_l \Gamma^i_{jk} \frac{dw^j}{dt} \frac{dw^k}{dt} \simeq -\frac{2}{9t} C^{im} C_{mnl} e^n t^{-1/3} = -\frac{1}{3t} \tensor{M}{^i_l} .  \end{equation}
The geodesic deviation equation then reads
\begin{equation} \frac{d^2 X^i}{dt^2} + \tensor{M}{^i_j} \frac{dX^j}{dt} - \frac{1}{3t} \tensor{M}{^i_j} X^j = 0. \end{equation}
The discriminant of this equation along an eigenvector of $\tensor{M}{^i_j}$ is readily computed in terms of an eigenvalue $\l$ of $\tensor{M}{^i_j}$: 
\begin{equation} \Delta = \l^2 + \frac{4}{3t} \l. \end{equation}
If $\l$ is less singular than $1/t$ as $t\to 0$, then we see that $\sqrt{\abs{\Delta}} \sim t^{-\b}$ for some $\b < 1$. Suppose the worst possible case -- that $\Delta$ is negative, signaling an oscillatory solution. By an analogue of the WKB approximation, it therefore follows that one must integrate up to a finite time (when $\frac{1}{1-\b}t^{1-\b} \approx \pi$) to possibly encounter a point where $X(t)$ returns to zero, so the geodesics will not re-intersect arbitrarily close to the singularity. Moreover, at such a finite time, we are already well outside the realm where our approximations in the limit $t \to 0$ can be trusted. 

It therefore follows that the only possible directions where we might expect the geodesics to intersect arbitrarily close to the singularity are along directions where the eigenvalues $\l$ of $\tensor{M}{^i_j}$ diverge as $1/t$ or worse. As we have shown previously, these directions are precisely the directions $v^i$ which lie in the kernel of $C_{ij}$. We further showed that the eigenvalues for each $v^i$ are given by $2/(3t)$ in all such directions. 

The resulting geodesic deviation equation then reads
\begin{equation} \frac{d^2 X}{dt^2} + \frac{2}{3t} \frac{dX}{dt} - \frac{2}{9t^2} X = 0. \end{equation}
This equation is readily solved as in Section \ref{sec:geo}. The result is 
\begin{equation} X(t) = C_1 t^{2/3} + C_2 t^{-1/3}. \end{equation}
We find exactly the result we deduced in Section \ref{sec:geo}, now in a much more general setting. The nonzero deviations correspond to moving around in different directions lying the kernel of $g_{ij}$ along which we may approach the singularity. In all cases, the nearby geodesics do not re-intersect arbitrarily close to $t = 0$, since the solution for $X(t)$ crosses $X(t) = 0$ at most once.

\bibliographystyle{JHEP}
\bibliography{bib.bib}

\providecommand{\href}[2]{#2}\begingroup\raggedright\begin{thebibliography}{10}

\bibitem{vafa_string_2005}
C.~Vafa, \emph{The {String} {Landscape} and the {Swampland}},  Oct., 2005.
\newblock 10.48550/arXiv.hep-th/0509212.

\bibitem{brennan_string_2018}
T.D.~Brennan, F.~Carta and C.~Vafa, \emph{The {String} {Landscape}, the {Swampland}, and the {Missing} {Corner}},  June, 2018.
\newblock 10.48550/arXiv.1711.00864.

\bibitem{van_beest_lectures_2022}
M.~van Beest, J.~Calderón-Infante, D.~Mirfendereski and I.~Valenzuela, \emph{Lectures on the {Swampland} {Program} in {String} {Compactifications}}, \href{https://doi.org/10.1016/j.physrep.2022.09.002}{\emph{Physics Reports} {\bfseries 989} (2022) 1}.

\bibitem{agmon_lectures_2023}
N.B.~Agmon, A.~Bedroya, M.J.~Kang and C.~Vafa, \emph{Lectures on the string landscape and the {Swampland}},  Mar., 2023.
\newblock 10.48550/arXiv.2212.06187.

\bibitem{palti_swampland_2019}
E.~Palti, \emph{The {Swampland}: {Introduction} and {Review}}, \href{https://doi.org/10.1002/prop.201900037}{\emph{Fortschritte der Physik} {\bfseries 67} (2019) 1900037}.

\bibitem{grana_swampland_2021}
M.~Graña and A.~Herráez, \emph{The {Swampland} {Conjectures}: {A} bridge from {Quantum} {Gravity} to {Particle} {Physics}},  July, 2021.
\newblock 10.48550/arXiv.2107.00087.

\bibitem{ooguri_geometry_2007}
H.~Ooguri and C.~Vafa, \emph{On the {Geometry} of the {String} {Landscape} and the {Swampland}}, \href{https://doi.org/10.1016/j.nuclphysb.2006.10.033}{\emph{Nuclear Physics B} {\bfseries 766} (2007) 21}.

\bibitem{trenner_asymptotic_2010}
T.~Trenner and P.M.H.~Wilson, \emph{Asymptotic curvature of moduli spaces for {Calabi}-{Yau} threefolds},  July, 2010.

\bibitem{marchesano_moduli_2023}
F.~Marchesano, L.~Melotti and L.~Paoloni, \emph{On the moduli space curvature at infinity},  Dec., 2023.

\bibitem{lu_hodge_2005}
Z.~Lu, \emph{On the {Hodge} {Metric} of the {Universal} {Deformation} {Space} of {Calabi}-{Yau} {Threefolds}},  May, 2005.

\bibitem{lu_curvature_2005}
Z.~Lu, \emph{On the {Curvature} {Tensor} of the {Hodge} {Metric} of {Moduli} {Space} of {Polarized} {Calabi}-{Yau} {Threefolds}},  June, 2005.

\bibitem{cecotti_moduli_2020}
S.~Cecotti, \emph{Moduli spaces of {Calabi}-{Yau} \$d\$-folds as gravitational-chiral instantons}, \href{https://doi.org/10.1007/JHEP12(2020)008}{\emph{Journal of High Energy Physics} {\bfseries 2020} (2020) 8}.

\bibitem{cecotti_special_2020}
S.~Cecotti, \emph{Special {Geometry} and the {Swampland}}, \href{https://doi.org/10.1007/JHEP09(2020)147}{\emph{Journal of High Energy Physics} {\bfseries 2020} (2020) 147}.

\bibitem{cecotti_swampland_2021}
S.~Cecotti, \emph{Swampland geometry and the gauge couplings}, \href{https://doi.org/10.1007/JHEP09(2021)136}{\emph{Journal of High Energy Physics} {\bfseries 2021} (2021) 136}.

\bibitem{mcnamara_cobordism_2019}
J.~McNamara and C.~Vafa, \emph{Cobordism {Classes} and the {Swampland}},  Oct., 2019.
\newblock 10.48550/arXiv.1909.10355.

\bibitem{liu_hodge_2016}
K.~Liu and Y.~Shen, \emph{Hodge metric completion of the moduli space of {Calabi}-{Yau} manifolds},  Sept., 2016.
\newblock 10.48550/arXiv.1305.0231.

\bibitem{banks_symmetries_2011}
T.~Banks and N.~Seiberg, \emph{Symmetries and {Strings} in {Field} {Theory} and {Gravity}}, \href{https://doi.org/10.1103/PhysRevD.83.084019}{\emph{Physical Review D} {\bfseries 83} (2011) 084019}.

\bibitem{polchinski_monopoles_2004}
J.~Polchinski, \emph{Monopoles, {Duality}, and {String} {Theory}}, \href{https://doi.org/10.1142/S0217751X0401866X}{\emph{International Journal of Modern Physics A} {\bfseries 19} (2004) 145}.

\bibitem{misner_classical_1957}
C.W.~Misner and J.A.~Wheeler, \emph{Classical physics as geometry}, \href{https://doi.org/10.1016/0003-4916(57)90049-0}{\emph{Annals of Physics} {\bfseries 2} (1957) 525}.

\bibitem{harlow_symmetries_2019}
D.~Harlow and H.~Ooguri, \emph{Symmetries in quantum field theory and quantum gravity},  June, 2019.
\newblock 10.48550/arXiv.1810.05338.

\bibitem{gaiotto_generalized_2015}
D.~Gaiotto, A.~Kapustin, N.~Seiberg and B.~Willett, \emph{Generalized {Global} {Symmetries}}, \href{https://doi.org/10.1007/JHEP02(2015)172}{\emph{Journal of High Energy Physics} {\bfseries 2015} (2015) 172}.

\bibitem{kirby_calculation_1990}
R.C.~Kirby and L.R.~Taylor, \emph{A calculation {ofPin}+ bordism groups}, \href{https://doi.org/10.1007/BF02566617}{\emph{Commentarii Mathematici Helvetici} {\bfseries 65} (1990) 434}.

\bibitem{giambalvo_langle_1971}
V.~Giambalvo, \emph{On \${\textbackslash}langle 8 {\textbackslash}rangle\$-cobordism}, \href{https://doi.org/10.1215/ijm/1256052508}{\emph{Illinois Journal of Mathematics} {\bfseries 15} (1971) 533}.

\bibitem{anderson_structure_1967}
D.W.~Anderson, E.H.~Brown and F.P.~Peterson, \emph{The {Structure} of the {Spin} {Cobordism} {Ring}}, \href{https://doi.org/10.2307/1970690}{\emph{Annals of Mathematics} {\bfseries 86} (1967) 271}.

\bibitem{dierigl_r7-branes_2023}
M.~Dierigl, J.J.~Heckman, M.~Montero and E.~Torres, \emph{R7-{Branes} as {Charge} {Conjugation} {Operators}}, .

\bibitem{dierigl_iib_2023}
M.~Dierigl, J.J.~Heckman, M.~Montero and E.~Torres, \emph{{IIB} string theory explored: {Reflection} 7-branes}, \href{https://doi.org/10.1103/PhysRevD.107.086015}{\emph{Phys. Rev. D} {\bfseries 107} (2023) 086015}.

\bibitem{basile_global_2024}
I.~Basile, A.~Debray, M.~Delgado and M.~Montero, \emph{Global anomalies \& bordism of non-supersymmetric strings}, \href{https://doi.org/10.1007/JHEP02(2024)092}{\emph{JHEP} {\bfseries 02} (2024) 092}.

\bibitem{garcia-etxebarria_dai-freed_2019}
I.~García-Etxebarria and M.~Montero, \emph{Dai-{Freed} anomalies in particle physics}, \href{https://doi.org/10.1007/JHEP08(2019)003}{\emph{Journal of High Energy Physics} {\bfseries 2019} (2019) 3}.

\bibitem{freed_reflection_2021}
D.S.~Freed and M.J.~Hopkins, \emph{Reflection positivity and invertible topological phases}, \href{https://doi.org/10.2140/gt.2021.25.1165}{\emph{Geometry \& Topology} {\bfseries 25} (2021) 1165}.

\bibitem{wan_higher_2019}
Z.~Wan and J.~Wang, \emph{Higher {Anomalies}, {Higher} {Symmetries}, and {Cobordisms} {I}: {Classification} of {Higher}-{Symmetry}-{Protected} {Topological} {States} and {Their} {Boundary} {Fermionic}/{Bosonic} {Anomalies} via a {Generalized} {Cobordism} {Theory}}, \href{https://doi.org/10.4310/AMSA.2019.v4.n2.a2}{\emph{Annals of Mathematical Sciences and Applications} {\bfseries 4} (2019) 107}.

\bibitem{debray_chronicles_2023}
A.~Debray, M.~Dierigl, J.J.~Heckman and M.~Montero, \emph{The {Chronicles} of {IIBordia}: {Dualities}, {Bordisms}, and the {Swampland}}, .

\bibitem{kaidi_non-supersymmetric_2023}
J.~Kaidi, K.~Ohmori, Y.~Tachikawa and K.~Yonekura, \emph{Non-supersymmetric heterotic branes},  Mar., 2023.

\bibitem{montero_cobordism_2021}
M.~Montero and C.~Vafa, \emph{Cobordism {Conjecture}, {Anomalies}, and the {String} {Lamppost} {Principle}}, \href{https://doi.org/10.1007/JHEP01(2021)063}{\emph{JHEP} {\bfseries 01} (2021) 063}.

\bibitem{friedrich_cobordism_2024}
B.~Friedrich, A.~Hebecker and J.~Walcher, \emph{Cobordism and bubbles of anything in the string landscape}, \href{https://doi.org/10.1007/JHEP02(2024)127}{\emph{JHEP} {\bfseries 02} (2024) 127}.

\bibitem{dierigl_swampland_2021}
M.~Dierigl and J.J.~Heckman, \emph{On the {Swampland} {Cobordism} {Conjecture} and {Non}-{Abelian} {Duality} {Groups}}, \href{https://doi.org/10.1103/PhysRevD.103.066006}{\emph{Physical Review D} {\bfseries 103} (2021) 066006}.

\bibitem{liu_global_2016}
K.~Liu, Y.~Shen and A.~Todorov, \emph{A {Global} {Torelli} {Theorem} for {Calabi}-{Yau} {Manifolds}},  Dec., 2016.
\newblock 10.48550/arXiv.1112.1163.

\bibitem{lu_weil-petersson_2005}
Z.~Lu and X.~Sun, \emph{Weil-{Petersson} geometry on moduli space of polarized {Calabi}-{Yau} manifolds},  Oct., 2005.

\bibitem{cartan_geometrie_1925}
E.~Cartan, \emph{La géométrie des espaces de {Riemann}}, .

\bibitem{hadamard_surfaces_1898}
J.~Hadamard, \emph{Les surfaces à courbures opposées et leurs lignes géodésiques}, {\emph{Journal de Mathématiques Pures et Appliquées} {\bfseries 4} (1898) 27}.

\bibitem{carroll_spacetime_2004}
S.M.~Carroll, \emph{Spacetime and {Geometry}: {An} {Introduction} to {General} {Relativity}}, Addison Wesley (2004).

\bibitem{cremmer_n8_1978}
E.~Cremmer and B.~Julia, \emph{The {N}=8 {Supergravity} {Theory}. 1. {The} {Lagrangian}}, \href{https://doi.org/10.1016/0370-2693(78)90303-9}{\emph{Phys. Lett. B} {\bfseries 80} (1978) 48}.

\bibitem{cremmer_so8_1979}
E.~Cremmer and B.~Julia, \emph{The {SO}(8) {Supergravity}}, \href{https://doi.org/10.1016/0550-3213(79)90331-6}{\emph{Nucl. Phys. B} {\bfseries 159} (1979) 141}.

\bibitem{julia_group_1981}
B.~Julia, \emph{Group disintegrations}, University Press, United Kingdom (1981).

\bibitem{de_wit_local_1985}
B.~de~Wit and H.~Nicolai, \emph{{LOCAL} {SU}(8) {INVARIANCE} {IN} d = 11 {SUPERGRAVITY}},  in \emph{Nuffield {Workshop} on {Supersymmetry} and its {Applications}}, Oct., 1985.

\bibitem{fre_general_2006}
P.~Fre', F.~Gargiulo, K.~Rulik and M.~Trigiante, \emph{The general pattern of {Kac} {Moody} extensions in supergravity and the issue of cosmic billiards}, \href{https://doi.org/10.1016/j.nuclphysb.2006.02.001}{\emph{Nuclear Physics B} {\bfseries 741} (2006) 42}.

\bibitem{eberlein_structure_1997}
P.B.~Eberlein, \emph{Structure of {Symmetric} {Spaces} of {Noncompact} {Type}},  in \emph{Geometry of {Nonpositively} {Curved} {Manifolds}}, Chicago {Lectures} in {Mathematics}, (Chicago, IL), pp.~66--169, University of Chicago Press (1997), \href{https://press.uchicago.edu/ucp/books/book/chicago/G/bo3633915.html}{https://press.uchicago.edu/ucp/books/book/chicago/G/bo3633915.html}.

\bibitem{narain_new_1986}
K.S.~Narain, \emph{New heterotic string theories in uncompactified dimensions {\textless} 10}, \href{https://doi.org/10.1016/0370-2693(86)90682-9}{\emph{Physics Letters B} {\bfseries 169} (1986) 41}.

\bibitem{narain_note_1987}
K.S.~Narain, M.H.~Sarmadi and E.~Witten, \emph{A {Note} on {Toroidal} {Compactification} of {Heterotic} {String} {Theory}}, \href{https://doi.org/10.1016/0550-3213(87)90001-0}{\emph{Nucl. Phys. B} {\bfseries 279} (1987) 369}.

\bibitem{griffiths_locally_1969}
P.~Griffiths and W.~Schmid, \emph{Locally homogeneous complex manifolds}, \href{https://doi.org/10.1007/BF02392390}{\emph{Acta Mathematica} {\bfseries 123} (1969) 253}.

\bibitem{lu_geometry_2005}
Z.~Lu, \emph{On the {Geometry} of {Classifying} {Spaces} and {Horizontal} {Slices}},  May, 2005.

\bibitem{milnor_curvatures_1976}
J.~Milnor, \emph{Curvatures of left invariant metrics on lie groups}, \href{https://doi.org/10.1016/S0001-8708(76)80002-3}{\emph{Advances in Mathematics} {\bfseries 21} (1976) 293}.

\bibitem{azencott_homogeneous_1976}
R.~Azencott and E.N.~Wilson, \emph{Homogeneous {Manifolds} with {Negative} {Curvature}. {I}}, \href{https://doi.org/10.2307/1999731}{\emph{Transactions of the American Mathematical Society} {\bfseries 215} (1976) 323}.

\bibitem{noauthor_topics_1984}
\emph{Topics in {Transcendental} {Algebraic} {Geometry}. ({AM}-106), {Volume} 106 {\textbar} {Princeton} {University} {Press}},  June, 1984.

\bibitem{mohsen_construction_2019}
J.-P.~Mohsen, \emph{Construction of negatively curved complete intersections},  Sept., 2019.

\bibitem{liu_curvatures_2016}
K.~Liu, X.~Sun, X.~Yang and S.-T.~Yau, \emph{Curvatures of moduli space of curves and applications},  Apr., 2016.
\newblock 10.48550/arXiv.1312.6932.

\bibitem{viaclovsky_kahler_2018}
J.A.~Viaclovsky, ``Ka¨hler manifolds, {Ricci} curvature, and hyperka¨hler metrics.'' June, 2018.

\bibitem{caorsi_geometric_2018}
M.~Caorsi and S.~Cecotti, \emph{Geometric classification of 4d \${\textbackslash}mathcal\{{N}\}=2\$ {SCFTs}}, \href{https://doi.org/10.1007/JHEP07(2018)138}{\emph{Journal of High Energy Physics} {\bfseries 2018} (2018) 138}.

\bibitem{candelas_pair_1991}
P.~Candelas, X.C.~De~La~Ossa, P.S.~Green and L.~Parkes, \emph{A pair of {Calabi}-{Yau} manifolds as an exactly soluble superconformal theory}, \href{https://doi.org/10.1016/0550-3213(91)90292-6}{\emph{Nuclear Physics B} {\bfseries 359} (1991) 21}.

\bibitem{blumenhagen_refined_2018}
R.~Blumenhagen, D.~Klaewer, L.~Schlechter and F.~Wolf, \emph{The refined {Swampland} {Distance} {Conjecture} in {Calabi}-{Yau} moduli spaces}, \href{https://doi.org/10.1007/JHEP06(2018)052}{\emph{Journal of High Energy Physics} {\bfseries 2018} (2018) 52}.

\bibitem{blumenhagen_emergent_2023}
R.~Blumenhagen, N.~Cribiori, A.~Gligovic and A.~Paraskevopoulou, \emph{The {Emergent} {M}-theory {Limit}}, .

\bibitem{eguchi_n2_1996}
T.~Eguchi and K.~Hori, \emph{\${N}=2\$ {Superconformal} {Field} {Theories} in \$4\$ {Dimensions} and {A}-{D}-{E} {Classification}},  July, 1996.

\bibitem{bharadwaj_approaching_2023}
S.~Bharadwaj and E.~D'Hoker, \emph{Approaching {Argyres}-{Douglas} theories},  Oct., 2023.
\newblock 10.48550/arXiv.2310.07703.

\bibitem{brandenberger_superstrings_1989}
R.H.~Brandenberger and C.~Vafa, \emph{Superstrings in the {Early} {Universe}}, \href{https://doi.org/10.1016/0550-3213(89)90037-0}{\emph{Nucl. Phys. B} {\bfseries 316} (1989) 391}.

\end{thebibliography}\endgroup

\end{document}